\documentclass{SciPost}

\hypersetup{
    colorlinks,
    linkcolor={red!50!black},
    citecolor={blue!50!black},
    urlcolor={blue!80!black}
}
\usepackage{textalpha}
\usepackage[section]{placeins}
\allowdisplaybreaks
\usepackage{xcolor}
\usepackage{array}

\usepackage{tikz}
\usepackage{tikz-feynman}
\tikzfeynmanset{compat=1.1.0}
\usepackage[bitstream-charter]{mathdesign}
\usepackage[bitstream-charter]{mathdesign}
\DeclareSymbolFont{usualmathcal}{OMS}{cmsy}{m}{n}
\DeclareSymbolFontAlphabet{\mathcal}{usualmathcal}

\fancypagestyle{SPstyle}{
\fancyhf{}
\lhead{\colorbox{scipostblue}{\bf \color{white} ~SciPost Physics }}
\rhead{{\bf \color{scipostdeepblue} ~Submission }}

\fancyfoot[C]{\textbf{\thepage}}
}

\begin{document}

\pagestyle{SPstyle}

\begin{center}{\Large \textbf{\color{scipostdeepblue}{
Amplitude mode in two-dimensional coherent spectroscopy of \\ weak-coupling antiferromagnets
}}}\end{center}

\begin{center}\textbf{
Jiyu Chen\textsuperscript{1,2,3},
Naoto Tsuji\textsuperscript{4,5,6} and
Philipp Werner\textsuperscript{3}
}\end{center}

\begin{center}
{\bf 1} Institute of Physics, Chinese Academy of Sciences, Beijing 100190, China
\\
{\bf 2} Songshan Lake Materials Laboratory, Dongguan, Guangdong 523808, China
\\
{\bf 3} Department of Physics, University of Fribourg, 1700 Fribourg, Switzerland
\\
{\bf 4} Department of Physics, University of Tokyo, Hongo, Tokyo 113-8656, Japan
\\
{\bf 5} RIKEN Center for Emergent Matter Science (CEMS), Wako 351-0198, Japan
\\
{\bf 6} Trans-scale Quantum Science Institute, University of Tokyo, Bunkyo-ku, Tokyo 113-8656, Japan
\\[\baselineskip]
\end{center}

\section*{\color{scipostdeepblue}{Abstract}}
\textbf{\boldmath{%
		Two-dimensional coherent spectroscopy (2DCS) provides insights into the nonlinear response of correlated lattice systems. We simulate multipulse excitations in the Hubbard model using nonequilibrium dynamical mean-field theory to extract the 2DCS signal of weak-coupling antiferromagnets with and without local potential disorder. 
		By comparing calculations with static and dynamic Hartree terms, and analyzing the waiting-time dependence of the signal, we identify the contribution of the collective amplitude mode to the spectroscopic features and the relevant underlying processes. 
		With broadband pulses, the rephasing and nonrephasing peaks at the gap energy are found to be of predominant amplitude mode character. 
		Using narrow-band pulses, we also demonstrate a strong enhancement of these amplitude mode-related signals at a pulse frequency of half the gap size. 
}}

\vspace{\baselineskip}

\noindent\textcolor{white!90!black}{%
\fbox{\parbox{0.975\linewidth}{%
\textcolor{white!40!black}{\begin{tabular}{lr}%
  \begin{minipage}{0.6\textwidth}%
    {\small Copyright attribution to authors. \newline
    This work is a submission to SciPost Physics. \newline
    License information to appear upon publication. \newline
    Publication information to appear upon publication.}
  \end{minipage} & \begin{minipage}{0.4\textwidth}
    {\small Received Date \newline Accepted Date \newline Published Date}%
  \end{minipage}
\end{tabular}}
}}
}

\vspace{10pt}
\noindent\rule{\textwidth}{1pt}
\tableofcontents
\noindent\rule{\textwidth}{1pt}
\vspace{10pt}

\section{Introduction}
Among various excitation modes, the collective processes giving rise to coherent oscillations of an order parameter are a unique property of correlated many-body systems in an ordered phase. They can provide crucial information on the symmetry, phase transition behavior, and hidden fluctuations of the system \cite{tsuji2024}. Examples include magnetic excitations (magnons \cite{lu2017,zhang2024} and amplitude modes of quantum antiferromagnets \cite{ruegg2008, jain2017, hong2017}), coherent
lattice excitations 
\cite{folpini2017},   
as well as charge excitations 
(e.g., Josephson plasmons \cite{zhang2023,liu2024,gomez2024}).
The collective amplitude (Higgs) excitation of superconductors \cite{pekker2015, shimano2020}, which does not usually couple to light in linear order \cite{tsuji2015,tsuji2016,schwarz2020}, has been studied through Raman scattering \cite{sooryakumar1980, grasset2018, grasset2019},
terahertz (THz) third harmonic generation \cite{matsunaga2014, matsunaga2017, chu2020}, and THz pump-probe spectroscopies \cite{matsunaga2013, katsumi2017}.

Two-dimensional coherent spectroscopy (2DCS), an experimental technique employing a sequence of laser pulses, allows to reveal the nonlinear response of target systems~\cite{mukamel1995,hamm2011}. Widely used in chemistry to study excitation pathways in molecules~\cite{hamm2011}, this method has been recently adopted in the condensed matter community to study nonlinear optical properties of quantum materials~\cite{liu2025}. {Examples include quantum spin liquids~\cite{wan2019,li2021,gao2023,li2023}, superconductors~\cite{gomez2024,gomez2025,fiore2025}, strange metals~\cite{chaudhuri2025}, correlated metals~\cite{barbalas2023} and  insulators~\cite{chen2025}.} Multipulse protocols with well-controlled phases and time delays can also be used to overcome the energy-time uncertainty when measuring nonequilibrium properties of materials~\cite{randi2017}, which is beyond the capabilities of conventional transient absorption spectroscopies~\cite{boschini2024}.

Recent THz 2DCS investigations of the conventional superconductor NbN \cite{katsumi2024} and the multi-gap superconductor MgB$_2$ \cite{katsumi2025}  
have revealed a characteristic resonance at a frequency corresponding to the superconducting gap.
This is in contrast to previous nonlinear THz spectroscopy studies \cite{matsunaga2014}, where the resonance was found at half of the gap size.
The origin of the latter resonance and its relation to the amplitude mode in superconductors has been discussed in the literature, both for clean and disordered systems \cite{cea2016, tsuji2016, matsunaga2017, jujo2018, murotani2019, silaev2019, tsuji2020, shimano2020, Seibold2021}.
Since the Higgs amplitude mode and quasiparticle excitations share the same energy gap, it would be useful if one could discriminate Higgs amplitude mode contributions from those of quasiparticles by 2DCS.

Besides superconductors, 2DCS studies of the  amplitude mode could provide useful insights into other ordered quantum phases, such as excitonic insulators~\cite{golez2020}, superfluid Bose gases~\cite{endres2012}, antiferromagnets and charge density waves~\cite{tsuji2024}, which can be described by a similar Ginzburg-Landau theory as superconductors~\cite{pekker2015}. In some models, for example, the antiferromagnetic solution for repulsive interactions can be mapped onto the $s$-wave superconducting solution of the attractively interacting model by the Shiba transformation~\cite{Shiba1972}, which interchanges spin and charge (by mapping particles to holes for spin down electrons). 
While the Higgs mode in superconductors is intimately related to the amplitude mode in antiferromagnets through this duality transformation, the coupling to gauge fields is rather different (since the Shiba transformation exchanges the role of electric fields with magnetic fields).
In an interaction-quench setup, the amplitude modes of the two systems would behave in the same way.

In this work, we explore the signatures of the collective amplitude mode in weak-coupling antiferromagnets described by the half-filled Hubbard model. 
The amplitude mode in this system corresponds to oscillations in the staggered magnetization. 
We will also comment on the similarities and differences between these 2DCS results for antiferromagnets and the previously discussed nonlinear responses of superconductors.

We employ a recently developed theoretical framework \cite{chen2025} based on nonequilibrium dynamical mean-field theory (DMFT) \cite{aoki2014} to simulate the real-time dynamics induced by multiple laser pulses, and extract the 2DCS signal from the measured current. This approach mimics the experimental protocol, and can in principle also be used to study multi-dimensional spectra of nonequilibrium states \cite{chen2025}, although in the present study we focus on the nonlinear response of equilibrium systems. By analyzing the oscillations of the signals as a function of the waiting time, and comparing simulations with and without feedback of the order parameter dynamics, we clarify the processes underlying the prominent 2DCS signals and identify the peaks with dominant contribution from the amplitude mode. We further study the effect of local potential disorder at the semiclassical level. 

The paper is organized as follows: Section \ref{sec:model} introduces the model and the DMFT technique used to treat 2DCS. Section \ref{sec:results} presents the analysis of the 2DCS signal of weak coupling antiferromangets with and without disorders, while Sec.~\ref{sec:conclusions} is a brief conclusion. 
 
\section{Model and method}
\label{sec:model}

\subsection{Model}

We consider the single-band Hubbard model with the Hamiltonian,
	\begin{equation}
			H=  -v_\text{hop}\sum_{\langle i j\rangle \sigma} c_{i\sigma}^{\dagger} c_{j \sigma}+\sum_iU n_{i \uparrow} n_{i\downarrow}-\sum_i (\mu-\epsilon_i) n_i. \label{eq_lattice}
	\end{equation}
	Here, $c_{i\sigma}^{\dagger}$ creates an electron with spin $\sigma$ on lattice site $i$, $n_{i \sigma}=c^\dagger_{i\sigma}c_{i\sigma}$ is the density operator, $U$ the on-site interaction, $\epsilon_i$ the local orbital energy, 
	and $\mu$ the chemical potential. In a clean, half-filled system, $\epsilon_i=0$ and $\mu=U/2$. 
    When including the effect of impurity scattering,
    we assume that the local orbital energy $\epsilon_i$ in Eq.~\eqref{eq_lattice} is normally distributed with variance $\langle \epsilon_i^2\rangle=\gamma^2$, where $\gamma$ is the disorder strength. 
    Local potential disorder is one of the representative types of disorders that has been widely studied \cite{jujo2018, murotani2019, silaev2019, tsuji2020, Seibold2021,li2024}. 
    Within the semiclassical treatment of weak disorder, the precise form of the impurity potential is not important, but what matters is the impurity scattering rate ($\sim \gamma^2 D(\varepsilon_F)$, with $D(\varepsilon_F$) the density of states at the Fermi level) \cite{AbrikosovBook}.

	We solve this model on the infinitely connected Bethe lattice (which is a tree-type graph where all vertices have the same number of neighbors) and define the renormalized hopping parameter $v=v_\text{hop}/\sqrt{Z}$ (with the connectivity $Z\rightarrow \infty$) \cite{georges1996}.
	The corresponding noninteracting density of states is semi-elliptical with bandwidth $W=4v$, and we use $v=1$ as the unit of energy ($\hbar/v$ as the unit of time). In the following, we set $\hbar=1$.
{
The Bethe lattice significantly simplifies the self-consistency loop in DMFT, because the hybridization function of the effective impurity problem can be determined directly from the local impurity Green's functions \cite{georges1996}, without an explicit calculation of lattice Green's functions. This saves memory and CPU time and enables efficient real-time simulations. Furthermore, the finite bandwidth of the Bethe lattice is more physical than, e.~g., the Gaussian density of states of the infinite-dimensional hypercubic lattice, and its bipartite nature makes it suitable for the study of antiferromagnets.
While a Bethe lattice with electric field may seem artificial, we can treat the effect of an electric field by assuming that half of the bonds are parallel to the field and half of them antiparallel. A detailed discussion can be found in  Ref.~\cite{werner2017}.}

The half-filled repulsive Hubbard model on the
Bethe lattice is antiferromagnetically ordered at low temperatures. Inside the antiferromagnetic (AFM) phase, there is a crossover from a weak-coupling (Slater-) antiferromagnet to a strong-coupling (Heisenberg-) antiferromagnet around $U\approx W$~\cite{pruschke2003}. 
	
\subsection{Methodology}

We use nonequilibrium DMFT~\cite{georges1996,aoki2014} to calculate the response of the Hubbard model to electric field pulses. 
DMFT maps the lattice model \eqref{eq_lattice} to a quantum impurity model, with a self-consistently determined bath. In the clean case ($\gamma=0$), this mapping is exact in the infinite-connectivity limit considered here. To study the nonequilibrium evolution, we introduce nonequilibrium Green's functions and solve the DMFT equations on the three-branch Kadanoff-Baym contour \cite{freericks2006,aoki2014}. 

\subsubsection {Weak-coupling impurity solver}

\begin{figure}[t]
    \centering
        \begin{tikzpicture}
  \begin{feynman}
    \vertex (base);
    \vertex[above=2.5cm of base] (fa0);
    \vertex[right=0.1cm of fa0] (fa) {(a)};

    \vertex[right=2cm of fa0] (fb) {(b)};
    \vertex[right=5.5cm of fa0] (fc) {(c)};

    \vertex[right=5mm of base] (a);
    
    \vertex[right=5mm of a] (b);
    \vertex[right=2mm of a] (e);
    \vertex[right=3mm of b] (f);

    \vertex[right=1cm of a] (c);
    \vertex[above=1cm of b] (d);
    \diagram* [thick]{
      (e) -- (f),
    };
    \vertex[above=1.5cm of a] (u1);
    \vertex[above=1.5cm of c] (u2);

    \diagram* {
      (u1) -- [double,double distance=0.2ex,thick,with arrow=0.5,arrow size=0.2em, half left, looseness=1.5] (u2)
            -- [double,double distance=0.2ex,thick, half left, looseness=1.5] (u1),
    };

    \diagram* {
      (b) -- [photon, out=90, in=-90,thick] (d),
    };

    \vertex[right=2cm of base] (a0);
    \vertex[above=1.5cm of a0] (a);

    \vertex[right=2.5cm of a] (b);

    \vertex[below=1.5cm of a] (u);
    \vertex[below=1.5cm of b] (v);
    \diagram* {
      (a) -- [fermion, out=75, in=105,thick] (b),
      (b) -- [fermion, out=-105, in=-75,thick] (a),
      (a) -- [boson,thick] (u),
      (b) -- [boson,thick] (v),
      (u) -- [fermion,thick] (v),

    };

    \vertex[right=5.cm of base] (t1);

    \vertex[right=3cm of t1] (t2);

    \vertex[right=1.5cm of t1] (v1);
    \vertex[above=2cm of v1] (v2);

    \diagram*{
      (t1) --[double,double distance=0.2ex,thick,with arrow=0.5,arrow size=0.2em](t2),
    };

    \draw[dashed,thick] (t1) -- (v2);
    \draw[dashed,thick] (t2) -- (v2);
  \end{feynman}
        \end{tikzpicture}
    \caption{{Feynman diagrams for different self-energies. (a) The Hartree self-energy $\Sigma_\sigma^{H}$. (b) The second-order self-energy $\Sigma^{(2)}$. (c) The disorder induced impurity scattering self-energy $\Sigma^{\text{dis}}$. Double lines with arrows represent the local Green's function $G(t,t')$, single lines with arrows represent the bath Green's functions $\mathcal{G}(t,t')$, wiggly lines represent the Hubbard interaction $U$ and dashed lines represent the impurity scattering.
    }
    }
    \label{fig:self-energy}
\end{figure}

We will focus on weakly correlated antiferromagnets with $U<W$, and perturbatively expand the impurity self-energy up to second order in $U$. This approximate impurity solver is sometimes called iterated perturbation theory (IPT) \cite{georges1996}. For a detailed discussion of different types of weak-coupling impurity solvers, we refer to Ref.~\cite{tsuji2013}.

The first-order Hartree term $\Sigma_\sigma^H$ is calculated with the instantaneous interacting (bold) Green's function $G$, 
\begin{equation}
	\Sigma_\sigma^H (t,t')\equiv  Un_{\bar{\sigma}}(t)\delta_\mathcal{C}(t,t')=-iU(t)G^<_{\bar{\sigma}}(t,t)\delta_\mathcal{C}(t,t').\label{eq:se1}
\end{equation} 
For the second-order contribution to  the self-energy, $\Sigma_\sigma^{(2)}$, we use the bare diagram~\cite{tsuji2013},
\begin{equation}
	\Sigma_\sigma^{(2)}\left(t, t^{\prime}\right)=U(t) U\left(t^{\prime}\right) \mathcal{G}_{0, \sigma}\left(t, t^{\prime}\right) \mathcal{G}_{0, \bar{\sigma}}\left(t^{\prime}, t\right) \mathcal{G}_{0, \bar{\sigma}}\left(t, t^{\prime}\right).\label{eq:se2}
\end{equation}
Here, $\mathcal{G}_0$ is the Weiss Green's function, which is related to the hybridization function $\Delta(t,t')$ of the impurity model by \cite{aoki2014}
\begin{equation}
\mathcal{G}_{0, \sigma}^{-1}(t,t')=i \partial_t+\mu-\Delta_\sigma(t,t'). \label{eq:weiss}
\end{equation}
This bare diagram has been shown to produce a more accurate nonequilibrium dynamics in the weak-correlation regime than the conserving approximation \cite{baym1961} with boldified second-order diagram \cite{eckstein2010,tsuji2013}.

Restricting ourselves to a semiclassical treatment 
(i.e., the self-consistent Born approximation)
of the local potential disorder,
we add the impurity scattering self-energy \cite{kemper2018,tsuji2020}
\begin{equation}
\Sigma^{\text{dis}}_\sigma(t,t')= \gamma^2G_\sigma(t,t').
\label{eq:disorder}
\end{equation}
In a diagrammatic language, Eq.~\eqref{eq:disorder} results in the summation of rainbow-type disorder diagrams, but neglects diagrams with crossing disorder lines. This treatment does not capture interference effects and hence localization physics. 
Since we directly simulate the real-time evolution (implementing a probe pulse and measuring the current) to extract nonlinear response signals, our calculation is equivalent to including ladder-type vertex corrections in nonlinear susceptibilities (see, e.g., Ref.~\citeonline{salvador2025}~and~\citeonline{tsuji2026}).
We focus on the weak disorder regime (where $\gamma$ is smaller than the antiferromagnetic gap size as well as the hopping amplitude), in which the self-consistent Born approximation is supposed to be valid. {Figure~\ref{fig:self-energy} shows the Feynman diagrams for the self-energy contributions defined in Eqs.~\eqref{eq:se1},~\eqref{eq:se2}~and~\eqref{eq:disorder}.}

\subsubsection{Self consistency}
Starting from the atomic Green's function, one obtains the hybridization function using the antiferromagnetic Bethe lattice self-consistency relation \cite{georges1996}
\begin{equation}
\Delta_\sigma(t,t') = v^*(t) G_{\bar{\sigma}}(t,t') v(t'),\label{eq_delta} 
\end{equation}
where $\bar\sigma$ denotes the spin opposite to $\sigma$. $v^*(t)$ is the complex conjugate of the hopping parameter $v(t)$ at time $t$. The time-dependent complex phases of the hopping parameters originate from  the external electric field and will be introduced in the following subsection.

Given the hybridization function, one next calculates the Weiss Green's function by solving Eq.~\eqref{eq:weiss}, with the hybridization function augmented by the impurity self-energy,
\begin{equation}
	(i \partial_t+\mu)\mathcal{G}_{0, \sigma}(t,t')-[(\Delta_\sigma+\Sigma^{\text{dis}}_\sigma)*\mathcal{G}_{0, \sigma}](t,t')= \delta(t,t').
    \label{eq:Weiss dyson}
\end{equation}
This choice provides impurity corrections in the second-order self-energy diagram $\Sigma_\sigma^{(2)}$ [Eq.~(\ref{eq:se2})].
If $\gamma$ is sufficiently small, the self-energy correction from the local potential disorder
in Eq.~(\ref{eq:Weiss dyson}) should not significantly alter the results.

After constructing the total self-energy according to Eqs.~\eqref{eq:se1}, \eqref{eq:se2} and \eqref{eq:disorder}, $\Sigma_\sigma=\Sigma^H_\sigma+\Sigma^{(2)}_\sigma+\Sigma^\text{dis}_\sigma$, 
the impurity Green's function is calculated by solving the impurity Dyson equation
\begin{equation}
	(i \partial_t+\mu)G_\sigma(t,t')-[(\Delta_\sigma+\Sigma_\sigma)*G_\sigma](t,t')= \delta(t,t').
\end{equation}
This defines the new hybridization function via Eq.~\eqref{eq_delta}. 

{In a full real-time simulation, all the three self-energies presented in Fig.~\ref{fig:self-energy} are dynamically updated at each time step, until a self-consistent solution is found. However, as discussed later, one can also implement an alternative scheme for the time evolution, where the time-local Hartree self-energy (Fig.~\ref{fig:self-energy}(a)) is kept fixed at its equilibrium value and only the other two self-energies (Fig.~\ref{fig:self-energy}(b)(c)) are dynamically updated. This artificial time evolution with static Hartree self-energy allows to suppress the amplitude mode excitations} \cite{tsuji2020}.
The calculations are implemented with the open-source simulation framework \verb|NESSi| \cite{schuler2020}.

\subsubsection{Electric fields}

In the presence of an electric field $\boldsymbol{E}(t)$, the hopping amplitude  $v_{ij}(t) = ve^{-i\phi_{ij}(t)}$ between the neighboring sites $i$ and $j$ is dressed by the Peierls phase $\phi_{ij}(t) = \boldsymbol{A}(t) \cdot\boldsymbol{r}_{ij}$, with $\boldsymbol{A}(t) = -\int_0^t  dt'\boldsymbol{E}(t')$ the vector potential and $\boldsymbol{r}_{ij}=\boldsymbol{r}_j-\boldsymbol{r}_i$ the displacement between the sites \cite{aoki2014}. In our Bethe lattice setup {with lattice constant $a=1$}, we assume that half of the hoppings are parallel to the electric field, $v_p(t) = ve^{iA(t)}/\sqrt{2}$, and the other half are anti-parallel, $v_a(t) = ve^{-iA(t)}/\sqrt{2}$, which is appropriate for the treatment of collinear pulses.
 The $\sqrt{2}$ factor is a convention, which assures that the modified self-consistency condition 
\begin{equation}
\Delta_\sigma(t,t') = \sum_{i=a,p} v_i^*(t) G_{\bar{\sigma}}(t,t') v_i(t'),\label{eq_delta2}
\end{equation}
yields the usual bandwidth of $4v$ in the absence of the field \cite{werner2017}. In this study, we assume that the magnetic-field components of the laser pulses are negligible (as is usually the case in experimentally realistic situations).

\begin{figure}[t]
	\centering
	\includegraphics[width=0.99\linewidth]{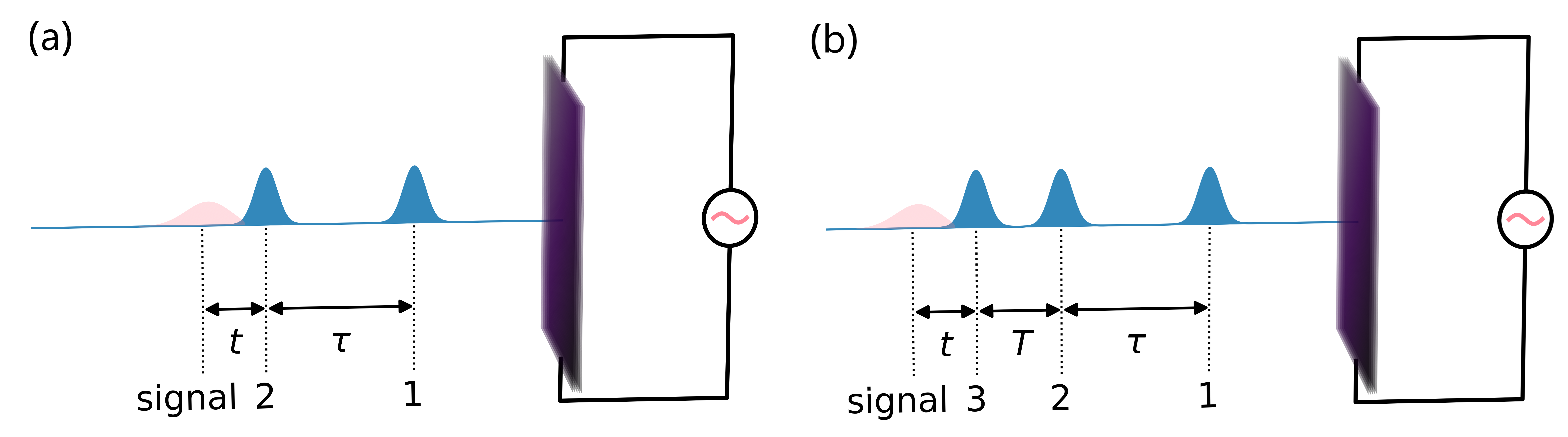}
	\caption{2DCS setup with (a) two and (b) three collinear pulse excitations (blue) and electric current measurement (red). 
	}\label{fig:1-exp}
\end{figure}

\subsubsection{Optical conductivity}

The current $\boldsymbol{j}$ induced by a weak electric field pulse $\boldsymbol{E}$ is determined by the optical conductivity $\sigma$ through the relation \cite{aoki2014}
\begin{equation}
	\boldsymbol{j}(t) = \int_0^t  dt'\sigma(t-t')\boldsymbol{E}(t').
\end{equation}
After Fourier transformation $\boldsymbol{j}(\omega) = \int_0^\infty d t \boldsymbol{j}(t) e^{i\omega t}$, this expression becomes $\boldsymbol{j}(\omega) = \sigma(\omega)\boldsymbol{E}(\omega)$.
In terms of the vector potential $\boldsymbol{A}(t) = -\int_0^t  dt'\boldsymbol{E}(t')$, 
one obtains the relation
\begin{equation}
	\sigma(\omega) = i\frac{{j}(\omega)}{\omega {A}(\omega)},\label{eq_cond}
\end{equation}
where we assumed that the current is parallel to the applied field.
If the pulse spectrum is broad enough, this equation allows to measure the conductivity.

\subsection{Multi-pulse setup}
In the 2DCS simulations, we perturb the system by two or three weak laser pulses, as shown in Fig.~\ref{fig:1-exp}. 
We consider a photocurrent spectroscopy setup, which combines collinear pump excitations with electric current detection \cite{chen2025}.
The first two pump pulses A and B are separated by a time delay $\tau$ (Fig.~\ref{fig:1-exp}(a)). In the three-pulse setup (Fig.~\ref{fig:1-exp}(b)), a third pulse C comes at time $T+\tau$, where $T$ is called the ``waiting time" .  The time delay between the last pump pulse and the photocurrent measurement is denoted by $t$. In the two-pulse scheme, the nonlinear current is calculated as $j_{NL}(t,\tau)\equiv j_\text{AB}(t+t_\tau)-j_\text{A}(t+t_\tau)-j_\text{B}(t+\tau)$. In the three-pulse scheme, $j_\text{NL}(t,\tau;T)\equiv j_\text{ABC}(t+\tau+T)-j_\text{AB}(t+\tau+T)-j_\text{AC}(t+\tau+T)-j_\text{BC}(t+\tau+T)$ $+j_\text{A}(t+\tau+T)+j_\text{B}(t+\tau+T)+j_\text{C}(t+\tau+T)$. 
The subscripts denote the presence of the pulses A, B or C. 
The induced transient states reveal information on excitation pathways, quantum coherences, and dephasing processes, without strong modifications of the population and density of states (DOS). 
Particularly, in the three-pulse setup, we are interested in the signal where each pump pulse interacts with the system once. In the two-pulse set-up, one of the pulses has to interact with the system twice to yield a third order response. By studying the dependence of the 2DCS signal on the waiting time $T$, one can detect the coherence of the intermediate states.

{For quantum systems with inversion symmetry, as we will study below, the second order response vanishes~\cite{branczyk2014}. Thus, both the two-pulse and the three-pulse protocols will generate third-order leading nonlinear excitations after subtraction of the linear contributions. However, as we will see later, they will still produce distinctly different results in some situations}. 

A subsequent Fourier transform of the variables ($\tau$,$t$) at fixed waiting time $T$ yields the two-dimensional intensity spectrum $|j_{NL}|(\omega_\tau,\omega_t)$ in the frequency plane. In the two-pulse setup (Fig.~\ref{fig:1-exp}(a)), we subtract only the single pulse signals (first order response), while the second-order response should be negligible due to inversion symmetry. In this setup, there is no waiting time, and we can directly calculate the Fourier transform with respect to ($\tau$,$t$).

\begin{figure}[t]
	\centering
	\includegraphics[width=0.7\linewidth]{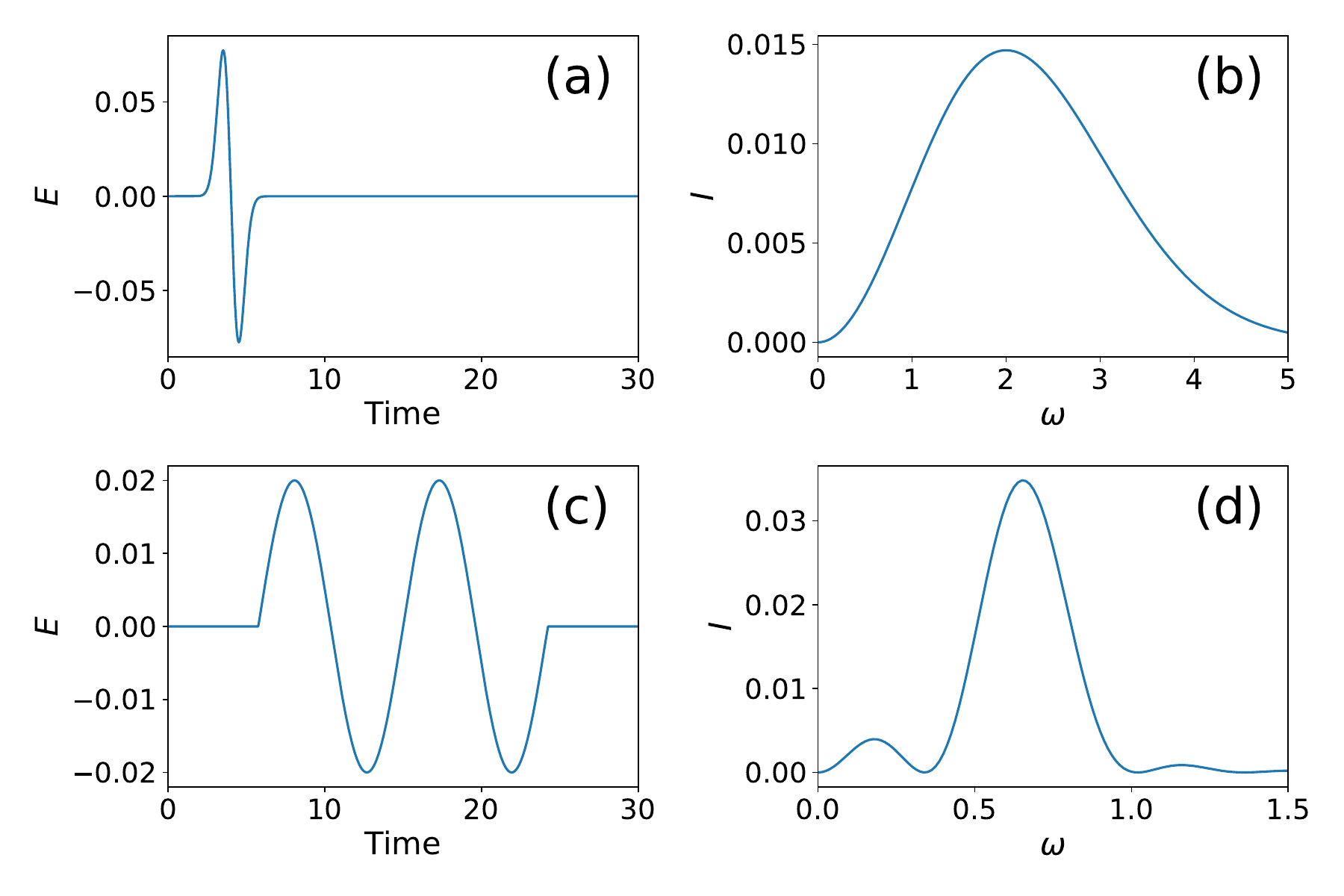}
	\caption{Electric field $E(t)$ (a)(c) and power spectrum $I(\omega) = |E(\omega)|^2$ 
	(b)(d) for the mono-cycle broadband pulse (a)(b) and a two-cycle ``monochromatic" pulse with frequency $\Omega=0.68$ (c)(d).
	}\label{fig:power} 	
\end{figure}

The pulses we consider are phase-stable monocycle and two-cycle laser pulses, as shown in Fig.~\ref{fig:power}. The broadband mono-cycle pulse centered at time $t_0$, with field strength $E(t)=$ $E_0 \frac{t-t_0}{\sqrt{2\pi}\sigma^2}\exp\Big(-\frac{(t-t_0)^2}{2\sigma^2}\Big)$,  has a Gaussian envelope with broadening $\sigma = 0.5$. The two-cycle pulse contains two sine waves with frequency $\Omega$. 
The pulse shapes $E(t)$ and power spectra $I(\omega)=|E(\omega)|^2$ are shown in panels (a,c) and (b,d), respectively. Similar pulses have been used in experimental studies~\cite{katsumi2024,katsumi2025}.

As we will see, the setups employing broadband pulses provide rich information on the possible excitations. Measurements with narrow band pulses yield additional insights into the excitation pathways associated with different signals, since for low enough pulse frequency, one needs multiple photons to excite across the gap. 

Note that we are interested here in studying the nonlinear response, typically third order response ($\propto E_0^3$), of the antiferromagnetically ordered equilibrium phase of the Hubbard model. Although our formalism is in principle valid for arbitrary field strengths, due to the non-perturbative treatment of the field, we consider here weak field pulses which do not significantly disturb the system.

\section{Results}
\label{sec:results}

\subsection{Equilibrium DOS and conductivity}

\begin{figure}[t]
	\centering
	\includegraphics[width=0.99\linewidth]{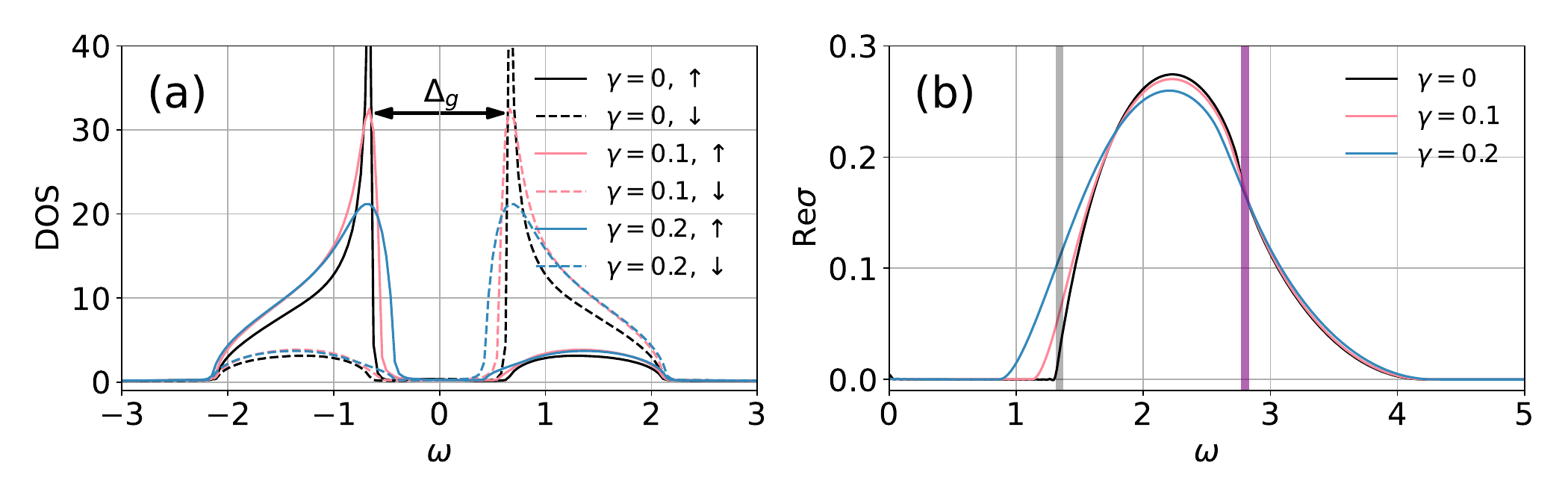}
	\caption{(a) Spin-resolved density of states in a system with $U=2$, $T=0.05$, and for the indicated disorder strengths $\gamma$. $\Delta_g$ is the full gap size of the clean system. (b) Real part of the optical conductivity $\sigma$. The grey and purple vertical lines correspond to $\Delta_g$ and $\Delta_g+\text{BW}$, where $\text{BW} \sim 1.4$ is the width of the upper/lower band.
	}\label{fig:dos_and_sigma}
\end{figure}

We perform simulations for a Hubbard model with $U=2$, which is in the weak-correlation regime that should be well described by the IPT solver. To be in the antiferromagnetic phase, we choose a low initial temperature $T=0.05$. For the DOS and optical conductivity calculation in Fig~\ref{fig:dos_and_sigma}, we simulate the time evolution up to 
$t_f = 160$ and use a time step $dt = 0.04$, corresponding to $N_t = 4000$ time points.
In the later 2DCS calculations, we simulate the time evolution up to 
$t_f = 100$ ($N_t = 2500$). No zero padding or additional windowing is used in the 2D Fourier transformations.

The calculated spin-resolved equilibrium spectral function is shown for different values of the disorder strength $\gamma$ in  Fig.~\ref{fig:dos_and_sigma}(a). The overall width of the DOS is similar to the noninteracting result ($W=4$). However, there is a gap of size $\Delta_g=1.35$, and the DOS exhibits sharp peaks at the gap edges. The occupied part of the DOS in Fig.~\ref{fig:dos_and_sigma} is dominated by $\sigma=\,\uparrow$, with the difference between the majority and minority occupation corresponding to the magnetization $m=n_\uparrow-n_\downarrow=0.6535$. 
On the neighboring sublattice, the spin polarization is opposite.

In Fig.~\ref{fig:dos_and_sigma}(b), we plot the real part of the optical conductivity, evaluated with a broadband electric field pulse and Eq.~\eqref{eq_cond}. The real part of the conductivity has a gap up to $\omega=\Delta_g$ and a maximum near $\omega=2.2$. (Note that we use here a notation which is different from some of the literature on superconductors, where the gap is defined as $2\Delta$.)
A rough understanding of the functional form of the conductivity can be obtained by considering the overlap between the empty (positive frequency) part of the spectrum and the occupied (negative frequency) part shifted by $\omega$. For $\omega\lesssim \Delta_g$, this overlap is approximately zero, then it grows rapidly up to $\omega\approx 2.2$, before decreasing again. The kink-like structure at $\omega\approx 2.8$ in the clean system (violet vertical line) marks the point where the shifted black spectrum in the left panel of Fig.~\ref{fig:dos_and_sigma} reaches the upper edge of the upper band.

The effect of local potential disorder is to broaden the sharp peaks in the DOS, which also leads to a narrowing of the gap. As a consequence, the real part of the conductivity becomes nonzero already for smaller values of~$\omega$.

\begin{figure}[t]
	\centering
	\includegraphics[width=0.6\linewidth]{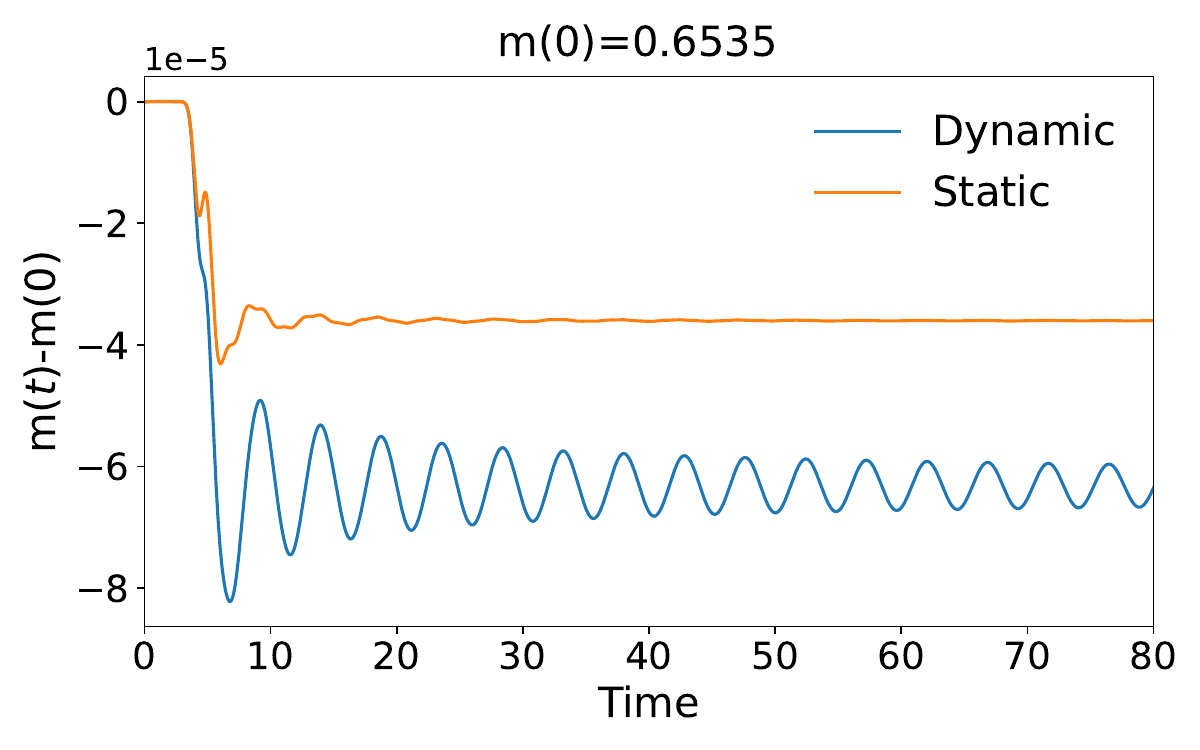}
	\caption{Time evolution of the magnetization $m$ after a short (broadband) excitation with $E_0=0.01$ and width $\sigma=0.5$. The orange curve shows the magnetization from a calculation where the Hartree term is fixed to its initial value.
    We plot the change of $m(t)$ relative to the initial value $m(0)=0.6535$.
    }\label{fig:m}
\end{figure}
\subsection{Evolution of the magnetization}

As shown by the blue line in Fig.~\ref{fig:m}, after a weak broadband excitation with amplitude $E_0=0.01$ and width $\sigma=0.5$, 
the magnetization is reduced relative to the initial value $m=0.6535$, but only very slightly. More importantly, the pulse induces magnetization oscillations (see blue curve) with frequency $\omega=1.35$, which corresponds to the gap size $\Delta_g$. This is the amplitude mode of the antiferromagnet, whose influence on the 2DCS signal is the focus of our study. In this full simulation with dynamic Hartree term, we update the Hartree term according to Eq.~\eqref{eq:se1}, which provides a feedback of the amplitude mode on the time-evolving state. For comparison, the orange curve in Fig.~\ref{fig:m} shows the result of a simulation performed with static Hartree term, where we keep the density in $\Sigma_\sigma^H$ (Eq.~\eqref{eq:se1}) fixed to the initial value $n_{\bar\sigma}(t=0)$. In the static Hartree calculation, the magnetization still changes after the pulse, because the occupations $G_\sigma^<(t,t)$ are modified, but it does not exhibit long-lived amplitude oscillations.
By simulating multi-pulse protocols with dynamic and static Hartree term, it is possible to quantify the effects of the amplitude mode on the 2DCS signal.

\subsection{2DCS signals}

\begin{figure}[t]
	\centering
    \includegraphics[width=0.95\linewidth]{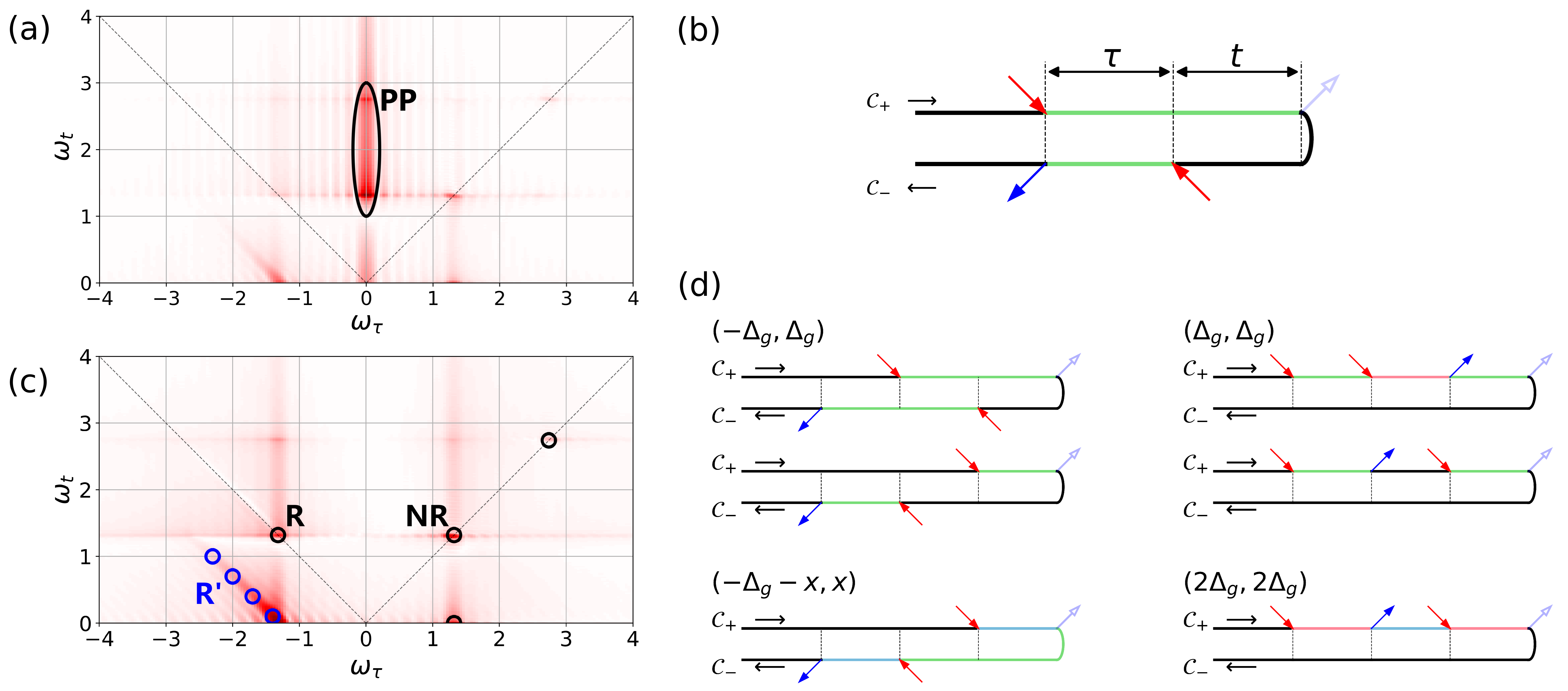}
	\caption{Two-pulse (a) and three-pulse (c) 2DCS signal obtained with weak broadband pulses. R and R' mark the rephasing signals. PP and NR marks pump-probe and nonrephasing signals, respectively.
	The three-pulse setup with fixed waiting time $T=10$ is used.
	The diagram in (b) illustrates the two-pulse PP process corresponding to the highlighted $\omega_\tau=0,\omega_t>0$ signals in panel (a). The first pulse interacts with both branches to create a population state (green). After the time delay $\tau$, the second pulse de-excites the system and generates a superposition state, which produces a current (open blue arrow) depending on time $t$.
    Panel (d) presents diagrams consistent with the waiting-time ($T$) dependence of the indicated $(\omega_\tau,\omega_t)$ signals. The color code of the segments indicates the ground state (black), and states with energy $\Delta_g$, $\Delta_g+x$, $2\Delta_g$ above the ground state (green, blue, pink).
    }\label{fig:broadband}
\end{figure}

\begin{table}[b]
\centering
\begin{tabular}{|p{3.4cm}|c|c|c|}
\hline
\textbf{\quad\quad\quad Location} & \textbf{\(E_0 = 0.01\)} & \textbf{\(E_0 = 0.02\) } & \textbf{Ratio} \\ \hline
\makebox[2cm][r]{$(-\Delta_g$}, \makebox[1cm][l]{~~$\Delta_g)$} & \(1.87823\times10^{-4}\) & \(1.49948\times10^{-3}\) & 7.98 \\ \hline
\makebox[2cm][r]{$(\Delta_g$}, \makebox[1cm][l]{~~$\Delta_g)$} & \(1.92373\times10^{-4}\) & \(1.54349\times10^{-3}\) & 8.02 \\ \hline
\makebox[2cm][r]{$(2\Delta_g$}, \makebox[1cm][l]{$2\Delta_g)$} & \(3.61789\times10^{-5}\) & \(2.93540\times10^{-4}\) & 8.11 \\ \hline
\makebox[2cm][r]{$(\Delta_g$}, \makebox[1cm][l]{$2\Delta_g)$} & \(3.00976\times10^{-5}\) & \(2.39780\times10^{-4}\) & 7.97 \\ \hline
\makebox[2cm][r]{$(-\Delta_g$}, \makebox[1cm][l]{$2\Delta_g)$} & \(4.96324\times10^{-5}\) & \(3.99727\times10^{-4}\) & 8.05 \\ \hline
\makebox[2cm][r]{$(-\Delta_g-0.4$}, \makebox[1cm][l]{~~$0.4)$} & \(1.71623\times10^{-4}\) & \(1.36728\times10^{-3}\) & 7.97 \\ \hline
\makebox[2cm][r]{$(-\Delta_g-0.1$}, \makebox[1cm][l]{~~$0.1)$} & \(2.87598\times10^{-5}\) & \(2.28999\times10^{-4}\) & 7.96 \\ \hline
\end{tabular}
\caption{Signal intensity at different encircled signals in Fig.~\ref{fig:broadband}(c) for pump intensities \(E_0 = 0.001\) and \(E_0 = 0.01\). Other parameters are the same as in Fig.~\ref{fig:broadband}.}\label{tab:tab1}
\end{table}

We first analyze the 2DCS spectra of the clean system ($\gamma=0$).  
Figure~\ref{fig:broadband} shows the spectra obtained with two (panel (a)) and three (panel (c)) pulses. 
The two-pulse measurement involves pump excitations at time $0$ and $\tau$, while the current is measured at time $\tau+t$. 
In panel (b), we illustrate the light-matter interaction sequence corresponding to the $\omega_\tau=0$ signal by a diagram representing the system's evolution along the Keldysh contour. Here, the lower (upper) branch of the contour depicts the time evolution of the bra (ket) of the
density matrix and the open blue arrow represents the current measurement. {The red (blue) arrows indicate the excitation (de-excitation) of the system when it interacts with light as it moves along the contour (in the direction determined by the Keldysh time ordering). The interpretation of these diagrams has been illustrated for simple examples in Ref.~\cite{chen2025}. In Fig.~\ref{fig:broadband}(b), the first laser interacts twice with both branches, i.e. the bra and ket, and generates a so-called polulation state $|e \rangle\langle e|$ during the time interval of length $\tau$.
Such two-pulse contributions are removed from the signal in the three-pulse setup, so that the corresponding 2DCS spectrum, shown in panel (c), is dominated by localized rephasing (R) and nonrephasing (NR) peaks at $(\omega_\tau,\omega_t)\approx (\pm \Delta_g, \Delta_g)$, and an elongated rephasing ridge (R$'$) extending along $(\omega_\tau,\omega_t) \approx (-\Delta_g - x, x)$ \cite{hamm2011}. Judging from the energies, the former may be associated with single-particle transitions across the gap or amplitude mode excitations, while the latter should be related to excitation and deexcitation processes within the subbands of the DOS. This is particularly evident on the rephasing side of the spectrum ($\omega_\tau<0$), where there is a band along $(-x-\Delta_g,x)$, up to an emission energy $\omega_t$ of approximately the width of the subbands in Fig.~\ref{fig:dos_and_sigma}(a). {Here $0<x<1.7$ denotes the energy of intra-band excitations.}
In Tab.~\ref{tab:tab1}, we present the intensity of the encircled signals in Fig.~\ref{fig:broadband}(c) using the pump strengths $E= 0.01$ and  $E= 0.02$.  Doubling the pump strength increases the signal by a factor of \(\sim 8\), confirming a third-order nonlinear response. 

\begin{figure}[t]
	\centering
	\includegraphics[width=0.99\linewidth]{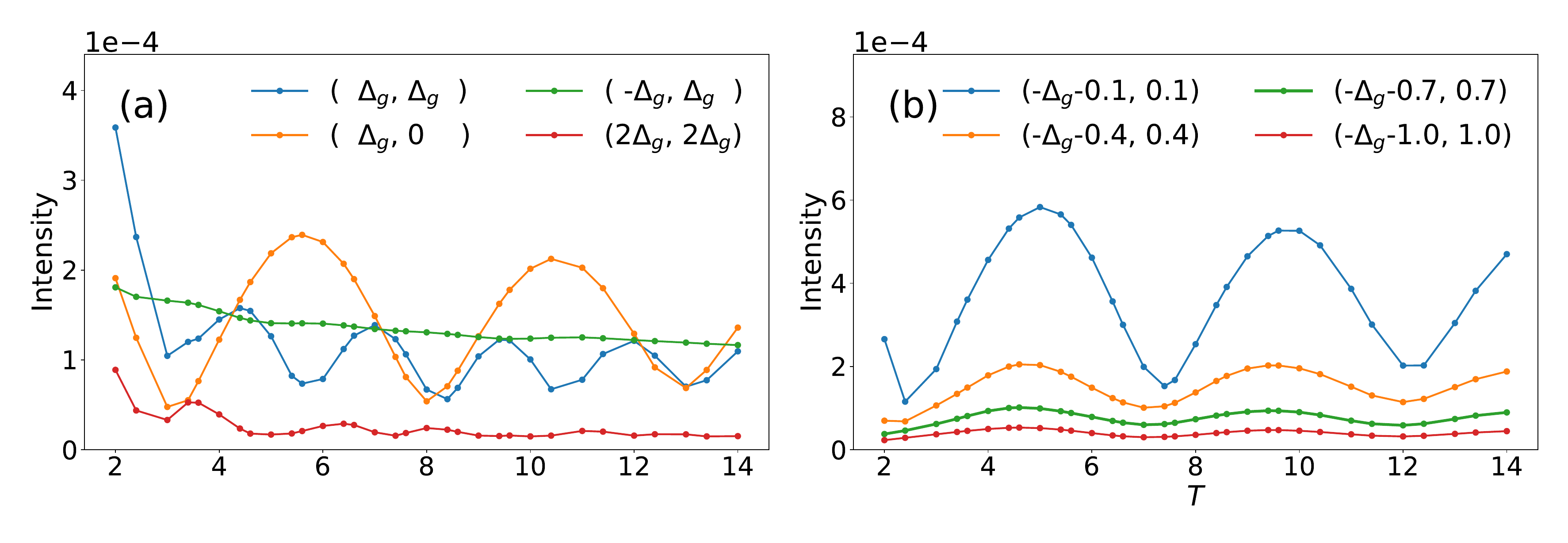}
	\caption{Signal intensities for the indicated $(\omega_\tau,\omega_t)$ as a function of the waiting time $T$. The signals corresponding to the black [blue] circles in Fig.~\ref{fig:broadband} are shown in panel (a) [(b)]. 
	}\label{fig:T}
\end{figure}

\begin{figure*}[t]
	\centering
	\includegraphics[width=1.0\linewidth]{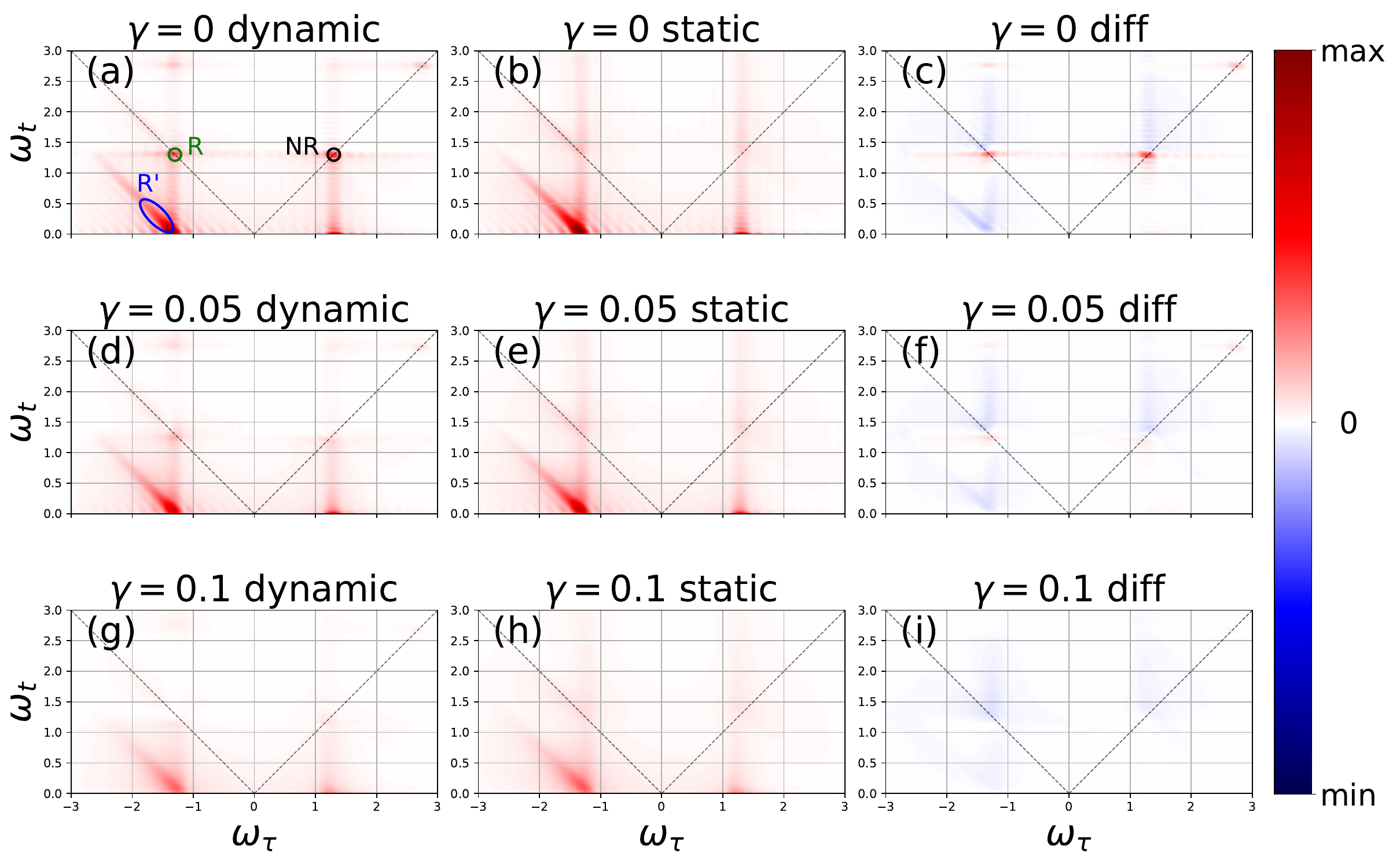}
	\caption{(a)(d)(g) 2DCS spectra for broadband monocycle excitations with indicated values of the local potential disorder $\gamma$. {The three-pulse setup with fixed waiting time $T=10$ is used.} In panel (a), the rephasing R, R', and nonrephasing NR signals are highlighted.
    (b)(e)(h) Analogous results obtained with a static Hartree term. (c)(f)(i) Difference 2DCS spectra $I_\text{diff}=I_\text{dynamic}-I_\text{static}$. The same color range is used in all the panels. 
	 }\label{fig:higgs}
\end{figure*}

We next analyze the oscillations of the relevant peaks in the three-pulse 2DCS signal. {As shown in Fig.~\ref{fig:T}(a), the intensity of the NR peak $(\Delta_g,\Delta_g)$ oscillates with frequency $2\Delta_g$, while the R signal $(-\Delta_g,\Delta_g)$ only shows a smooth decay, 
consistent with the results of previous model studies \cite{branczyk2014,chen2025}. The evolution of the R signal as a function of $T$ is consistent with its echo nature. Inelastic scattering processes, for example associated with excitations to the quasiparticle continuum, lead to cancellations.
Fitting the decay rate of this signal allows to quantify the life-time of the amplitude mode excitation}.
In Fig.~\ref{fig:T}(b), the series of R' signals along $(\omega_\tau,\omega_t)=(-x-\Delta_g,x)$ oscillates at frequency $\Omega=\Delta_g$, with an amplitude that increases as $x\rightarrow 0$. This could be explained for example by a superposition state $|g \rangle\langle g+e|$ or a state $|g \rangle\langle g+\text{Amp}|$, 
with ``$g$" denoting the ground state, ``$e$" a charge excitation across the gap (with energy $\Delta_g$), and ``Amp" denoting the amplitude mode (also with energy $\Delta_g$). Figure \ref{fig:T}(a) reveals that the peak located at $(\Delta_g,\Delta_g)$ oscillates at frequency $2.7$, which is {\it twice} the gap. This oscillation could come from a superposition state $|g\rangle\langle g+2\text{Amp}|$ or $|g\rangle\langle g+2e|$. In contrast, the peak intensity at $(-\Delta_g,\Delta_g)$ only shows a smooth decay. This behavior is expected for a population state of the type $|g\rangle\langle g|$, $|g+e\rangle\langle g+e|$ or $|g+\text{Amp}\rangle\langle g+\text{Amp}|$.

Diagrams consistent with the above-mentioned observations are sketched in Fig.~\ref{fig:broadband}(d). Here, black segments correspond to the ground state, green segments to states with energy $\Delta_g$ above the ground state, blue segments to states with energy $\Delta_g+x$ and pink segments to states with energy $2\Delta_g$.

\begin{figure*}[t]
		\centering
\includegraphics[width=0.99\linewidth]{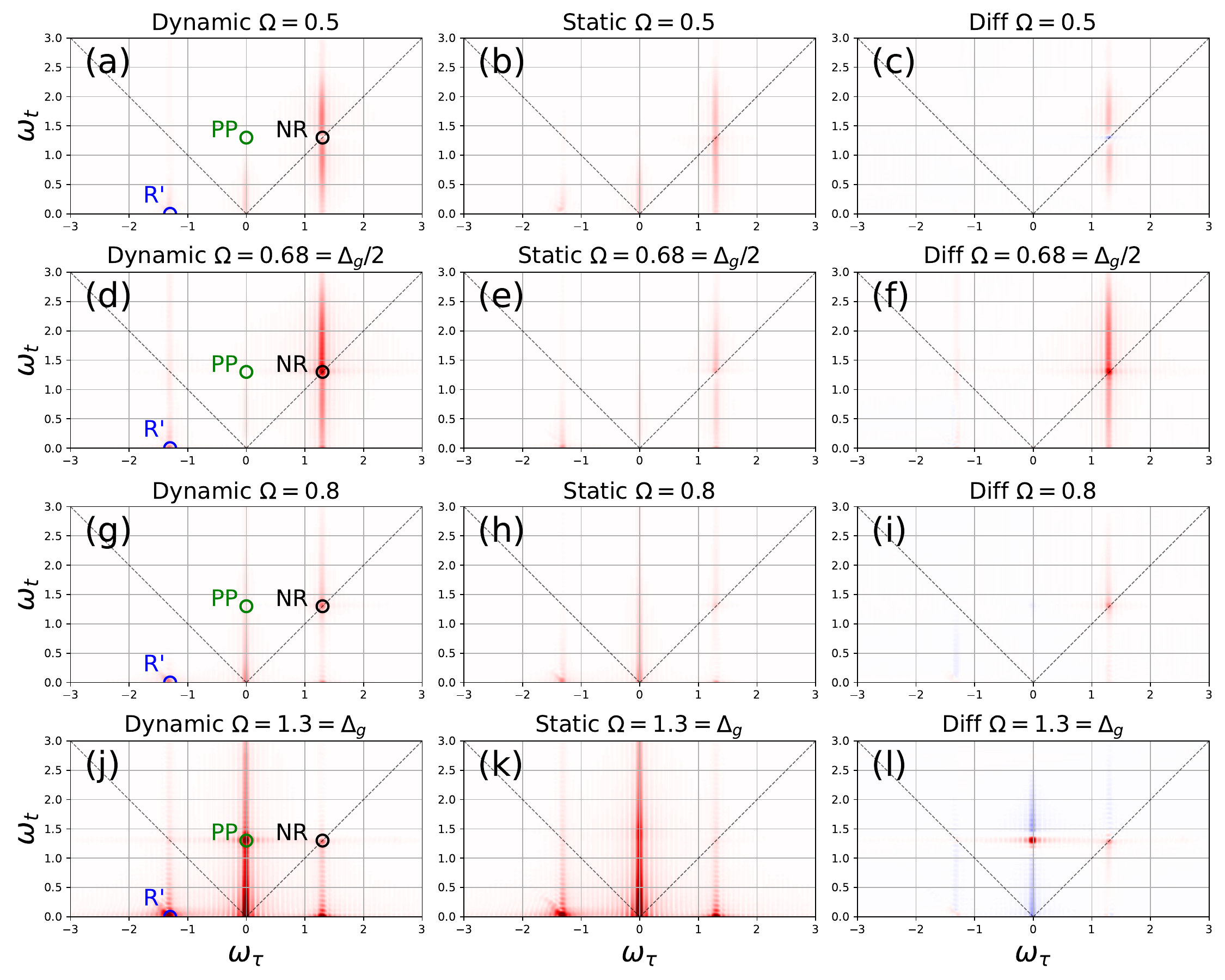}
	\caption{2DCS spectra of the clean system, obtained with two-cycle monochromatic pulses with the indicated frequencies $\Omega$, using the two-pulse setup. 
	The three columns correspond to the calculation with dynamical Hartree term (a)(d)(g)(i), static Hartree term (b)(e)(h)(k) and the difference spectrum (c)(f)(i)(l). In panel (a)(d)(g)(j), the NR, PP and R' signals are highlighted.
	}\label{fig:two_cycle2}
\end{figure*}

\subsection{Amplitude mode contribution and disorder effect}

In Fig.~\ref{fig:higgs}, we compare the 2DCS signal for three broadband pulses (left panels (a,d,g)) to the corresponding signal obtained with a static Hartree self-energy implementation (middle panels (b,e,h)). In the clean case ($\gamma=0$, first row), we notice that the prominent peaks at $(\pm\Delta_g,\Delta_g)$ almost disappear in the static Hartree calculation. 
In panel (c), we plot the difference spectrum for the clean system, defined as the difference between the absolute values of the 2DCS spectra with and without feedback from the self-consistently updated Hartree term. Here, the red color indicates signals contributed by the oscillations of the Hartree term, i.e., by the amplitude mode. It becomes clear from this analysis that the peaks at $(\pm \Delta_g,\Delta_g)$, whose intensity scales cubically with the field amplitude, are dominated by processes involving amplitude mode excitations.
 In contrast, the R' signals do not change significantly between the dynamic and static Hartree calculations, which shows that {amplitude mode} excitations play no significant role here. 

Combining this information with the observation in Fig.~\ref{fig:T}(a), we conclude that the dominant nonrephasing peak at $(\Delta_g,\Delta_g)$, which oscillates at twice the gap energy, originates from a superposition state with excited amplitude mode, like $|g\rangle\langle g+2\text{Amp}|$, and that the rephasing peak at $(-\Delta_g,\Delta_g)$ is related to the population state $|g+\text{Amp}\rangle\langle g+\text{Amp}|$. The R' signals, on the other hand, are dominated by superposition states of the type $|g \rangle\langle e|$.

The effect of local potential disorder is illustrated in the second and third row of Fig.~\ref{fig:higgs}. In the case of superconductors, it has been shown that the amplitude contribution to the third-order response becomes more prominent in the presence of {disorder}  \cite{jujo2018,murotani2019,silaev2019,tsuji2020}. Here, we find the opposite effect: while the 2DCS signals broaden due to the disorder scattering, the relative weight of the amplitude contribution to the $(\pm \Delta_g,\Delta_g)$ peaks decreases. 
We note that from a theoretical point of view, antiferromagnets and superconductors are not necessarily expected to react in the same way to {disorder}. In superconductors, the Anderson theorem~\cite{anderson1959} guarantees that the order parameter is robust against (nonmagnetic) impurities. This is not the case for the antiferromagnet, at least within our semiclassical description of the local potential disorder, where $m$ decreases from 0.653 to 0.650 and 0.638 when $\gamma$ increases from 0 to 0.1 and 0.2.

Let us comment on the relation to the recent THz 2DCS experiment for NbN superconductors \cite{katsumi2024}. While the observed enhancement of the signal at the gap frequency is similar to our case, there are several differences. First, we utilize broad-band pulse excitations in this section, while narrow-band pulses were used in the experiment. We will explore the difference between broad-band and narrow-band pulses in the next section. Second, as mentioned above, the amplitude mode-dominated signal at the gap frequency is fragile against {disorder} in the antiferromagnetic phase. This is in contrast to the case of superconductors, where the resonance peak at the gap energy survives even in the presence of strong {disorder}.
Another difference is that
we focus on peaks in the 2DCS spectra with $(\omega_\tau, \omega_t)=(\pm\Delta_g, \Delta_g)$, while the experiments measure the first-harmonic components in the nonlinear current intensity. We will also reexamine this point in the next section.

\subsection{ ``Monochromatic" excitations}
\subsubsection{ Two-pulse protocol}
Previous theoretical works have demonstrated a resonant enhancement of third-harmonic generation due to the amplitude mode at the pump frequency $\Omega=\Delta_g/2$ in superconductors \cite{tsuji2015,matsunaga2017}. It is thus interesting to study the amplitude mode-related 2DCS signal at $(\omega_\tau,\omega_t)=(\Delta_g,\Delta_g)$ for narrow-band two-cycle pulses with different frequencies. 
Figure~\ref{fig:two_cycle2} shows the 2DCS spectra obtained with the two-pulse protocol (Fig.~\ref{fig:1-exp}(a)) in the clean system, for pulse frequencies $\Omega=0.5<\Delta_g/2$ (a)(b)(c), $\Omega=0.68=\Delta_g/2$ (d)(e)(f), $\Delta_g>\Omega=0.80>\Delta_g/2$ (g)(h)(i), and $\Omega=\Delta_g$ (j)(k)(l).

\begin{figure}[t]
	\centering
	\includegraphics[width=0.75\linewidth]{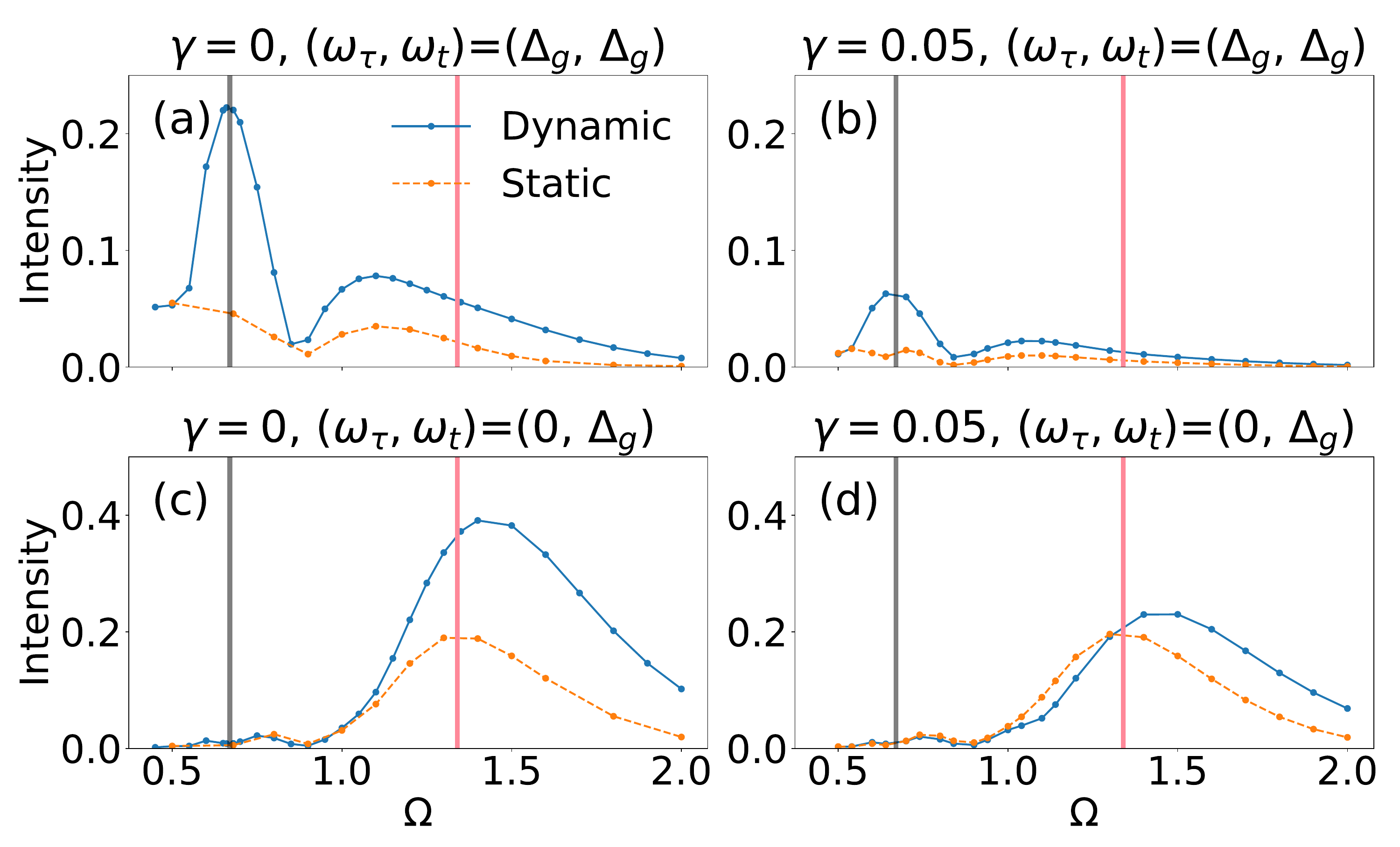}
	\caption{Intensity of the peaks at $(\omega_\tau,\omega_t)=(\Delta_g,\Delta_g)$ (top) and  $(\omega_\tau,\omega_t)=(0,\Delta_g)$ (bottom) for the clean (a)(c) and disordered ((b)(d), $\gamma=0.05$) system, plotted as a function of the pulse frequency $\Omega$. The 2DCS signal has been calculated with two-cycle monochromatic pulses and the two-pulse excitation protocol. The black (red) vertical line indicates the pulse frequency $\Delta_g/2$ ($\Delta_g$).
	}\label{fig:two_cycle_cut}
\end{figure}

\begin{figure}[t]
	\centering
	\includegraphics[width=0.77\linewidth]{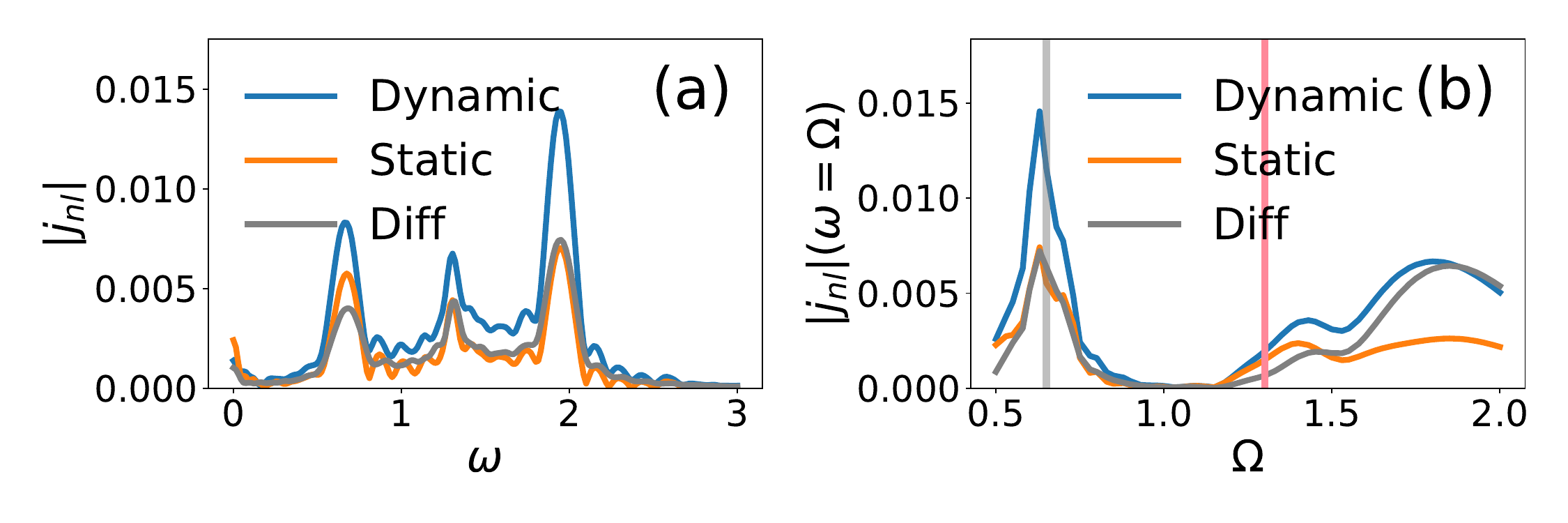}
	\caption{(a) Power spectrum of the nonlinear current of the clean system, excited with narrow-band pulses with frequency $\Omega=\Delta_g/2$. (b) First harmonic signal intensity as a function of the narrow-band excitation frequency $\Omega$. {The nonlinear current difference $|j_\text{diff}|=|j_\text{dynamic}-j_\text{static}|$ is measured in the time domain and then Fourier transformed to the frequency domain.}
    The black (red) vertical line indicates the pulse frequency $\Delta_g/2$ ($\Delta_g$).
	}\label{fig:nonlinear}
\end{figure}

\begin{figure*}[t]
	\centering
	\includegraphics[width=0.99\linewidth]{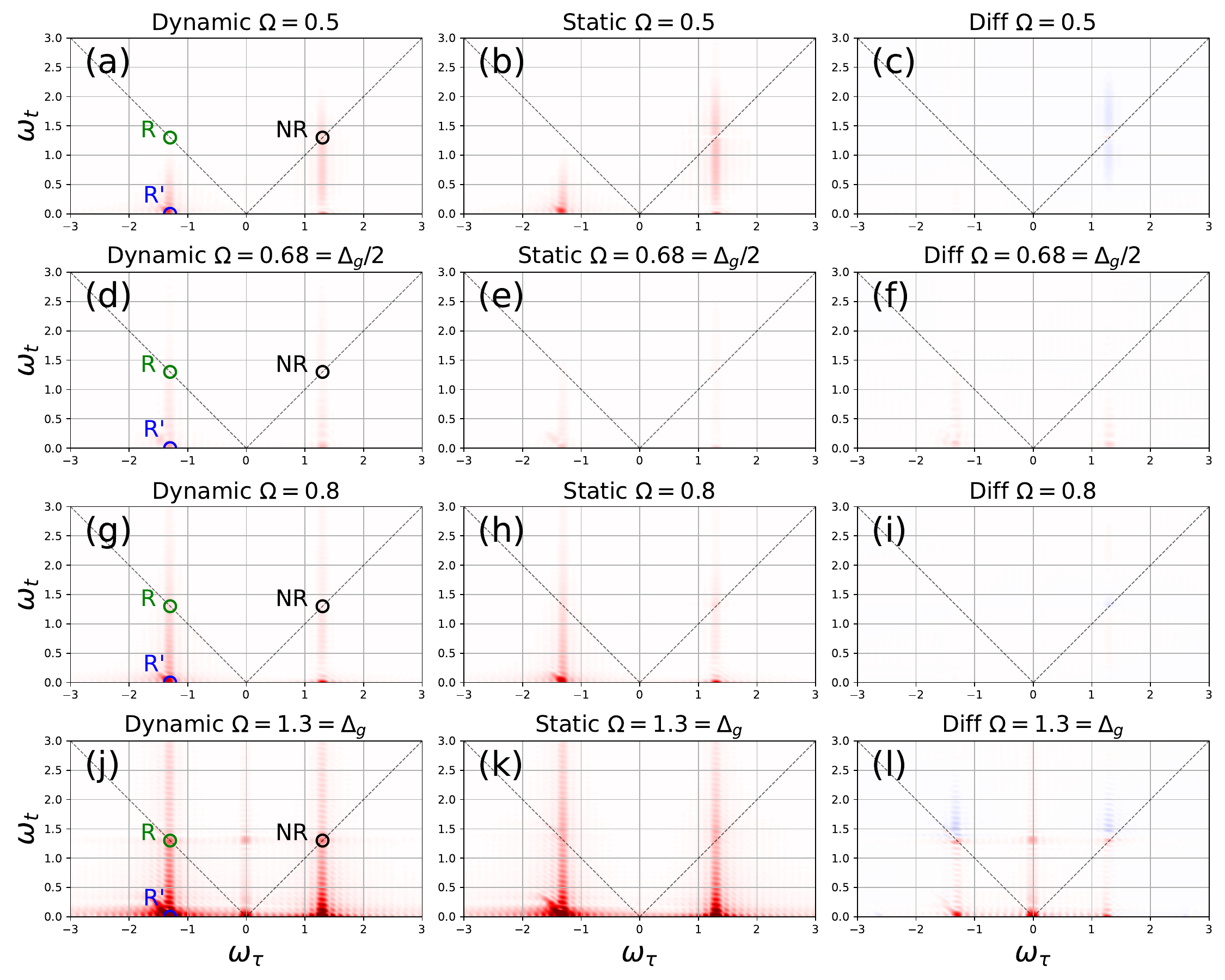}
	\caption{2DCS spectra of the clean system, obtained with two-cycle monochromatic pulses with the indicated frequencies $\Omega$, using the three-pulse setup. The three columns correspond to the calculation with dynamical Hartree term (a)(d)(g)(j), static Hartree term (b)(e)(h)(k) and the difference spectrum (c)(f)(i)(l). In panel (a)(d)(g)(j), NR, R and R' signals are highlighted.}\label{fig:two_cycle3}
\end{figure*}

Clearly, the nonrephasing signal at $(\omega_\tau,\omega_t)=(\Delta_g,\Delta_g)$ is enhanced near $\Omega=\Delta_g/2=0.68$, but only in the calculation with the dynamic Hartree term (left column). The intensity of this signal scales like $E_0^3$ (when increasing the field strength from $E_0=0.002$ to $E_0=0.02$, the signal increases from $4.8\times 10^{-6}$ to $4.9\times10^{-3}$).
This is consistent with a strong contribution from the resonantly excited amplitude mode. During the first pump, two photons with energy $\Delta_g/2$ are absorbed on one branch of the Keldysh contour, creating the amplitude mode excitation, while the second pump interacts once and produces a low-energy intraband excitation. 

When the pump frequency increases to $\Omega\simeq\Delta_g$, a single photon absorption is enough to produce an amplitude mode excitation, resulting in a configuration $|g+\text{Amp}\rangle\langle g+\text{Amp}|$ after the first pump (analogous to the two-pulse ``pump-probe" (PP) diagram shown in Fig.~\ref{fig:broadband}(b)).
During the time delay $\tau$ in the two-pulse measurement, the system is hence in this population state, and does not oscillate. This results in a large signal at $(\omega_\tau,\omega_t)=(0,\Delta_g)$ for $\Omega\gtrsim \Delta_g$, as seen in the bottom row of Fig.~\ref{fig:two_cycle2}.

In Fig.~\ref{fig:two_cycle_cut}, we plot the peak intensities at 
$(\omega_\tau,\omega_t)=(\Delta_g,\Delta_g)$ and $(0,\Delta_g)$ as a function of the pulse frequency $\Omega$, both for the clean system (left panels), and a disordered system with $\gamma=0.05$ (right panels).
In panels (a) and (b), we clearly see the enhancement of the 2DCS signal at $(\omega_\tau,\omega_t)=(\Delta_g,\Delta_g)$ near $\Omega=\Delta_g/2$,
which comes from the amplitude mode (compare the dynamic and static Hartree results). This enhancement is prominent both in the system without and with {disorder}, although in the latter case, the intensity of the signals is strongly reduced.
Panels (c) and (d) demonstrate the increase of the signal at $(\omega_\tau,\omega_t)=(0,\Delta_g)$ once the pulse frequency $\Omega$ becomes of the order of $\Delta_g$. In this case, the amplitude mode contribution dominates in the clean system, consistent with Fig.~\ref{fig:two_cycle2}(l), but becomes less significant in the presence of {disorder}.

In order to connect to the 2DCS experiments on superconductors with narrow-band pulses \cite{katsumi2024, katsumi2025},
we also calculate the power spectrum of the nonlinear current using two narrow-band pulse excitations with time delay $\tau=0$ and frequency $\Omega=\Delta_g/2$. As shown in Fig.~\ref{fig:nonlinear}(a), one can see that the power spectrum of the clean system exhibits three peaks at $\omega=\Delta_g/2$, $\Delta_g$ and $3\Delta_g/2$, corresponding to the first harmonic, second harmonic and third harmonic response , respectively. Note that in our inversion symmetric system, the second order response $\chi^{2}$ vanishes. The first, second and third harmonic response corresponds to the third-order nonlinear optical susceptibility $\chi^{(3)}(\Omega; +\Omega, -\Omega, +\Omega)$, $\chi^{(3)}(2\Omega; +\Omega, \Omega, 0)$ and $\chi^{(3)}(3\Omega; +\Omega, +\Omega, +\Omega)$, respectively \cite{katsumi2024,katsumi2025}.

In Fig.~\ref{fig:nonlinear}(b), we plot the first harmonic response as a function of frequency $\Omega$ of the narrow-band excitations. 
This component 
is not easy to measure in conventional pump-probe spectroscopy, but becomes accessible through 2DCS.
We observe a pronounced peak in the first harmonic component at $\Omega\approx\Delta_g/2$ and a continuum at $\Omega\gtrsim\Delta_g$. 
The former is significantly enhanced by the amplitude mode contribution, which can be simply understood from the two-photon absorption process (as in the case of third harmonic generation \cite{tsuji2015}).
This observation has to be contrasted with the experiment on NbN superconductors \cite{katsumi2024}, where a clear resonance is found at the gap frequency (not at half of the gap). The origin of this qualitative difference is at present unclear, but an obvious fact is that here we study the antiferromagnetic phase rather than the superconducting phase.
Analyzing the 2DCS signals of superconductors will be left for a future investigation.

\begin{figure}[t]
	\centering
	\includegraphics[width=0.7\linewidth]{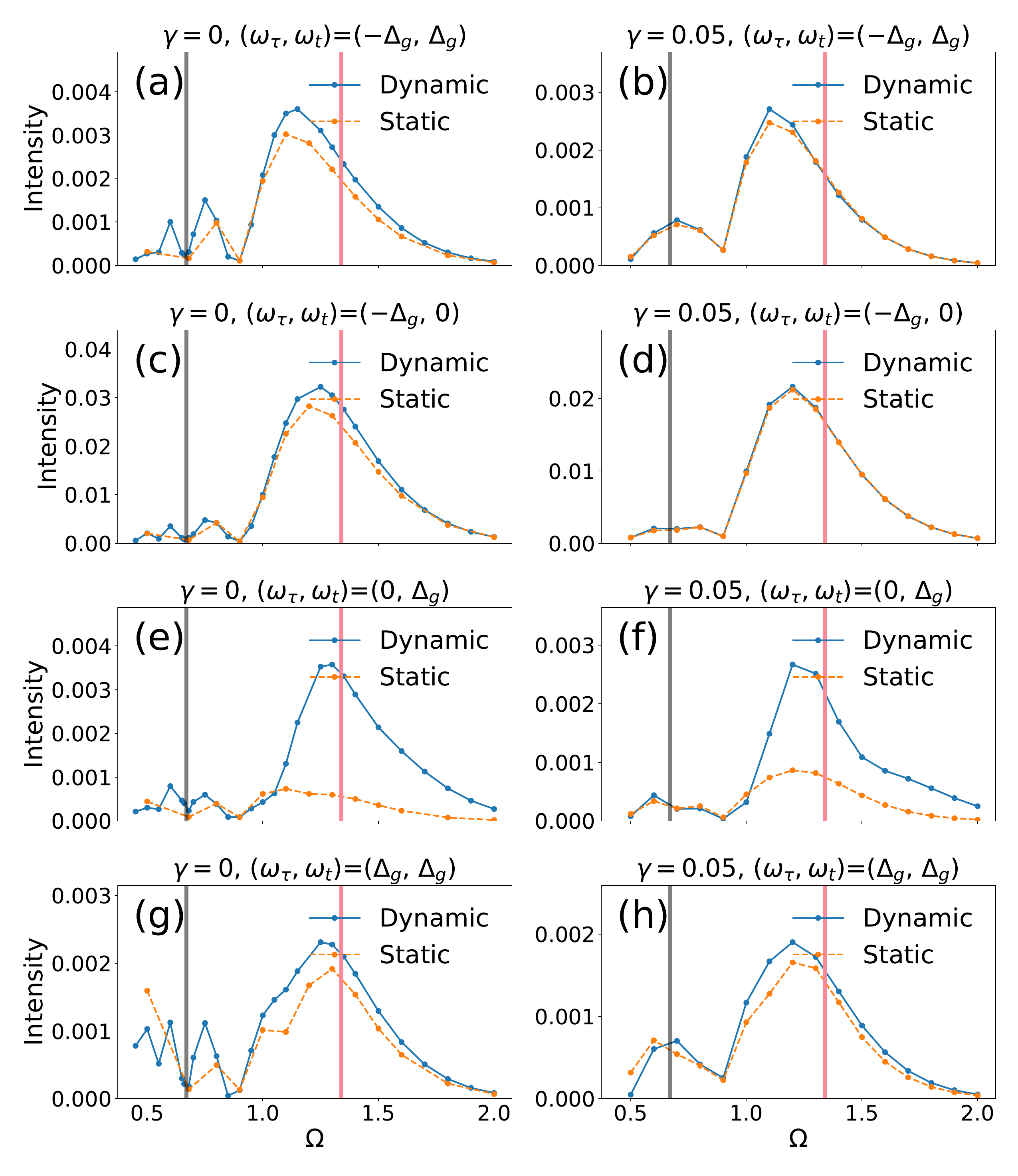}
	\caption{Intensity of the 2DCS signal at the indicated $(\omega_\tau,\omega_t)$  for the clean (a)(c)(e)(g) and disordered ((b)(d)(f)(h), $\gamma=0.05$) system, plotted as a function of the pulse frequency $\Omega$. The 2DCS spectra have been calculated with two-cycle monochromatic pulses and the three-pulse excitation protocol. The black (red) vertical line indicates the pulse frequency $\Delta_g/2$ ($\Delta_g$).
	}\label{fig:two_cycle_cut3}
\end{figure}

\subsubsection{ Three-pulse protocol}
If we use the three-pulse protocol, the 2DCS signal at $(\omega_\tau,\omega_t)=(\Delta_g,\Delta_g)$ shows no enhancement near pump frequency $\Omega=\Delta_g/2$, see Figs.~\ref{fig:two_cycle3} and \ref{fig:two_cycle_cut3}. At this energy, the third-order signal is very weak, since single photons can neither create {amplitude mode} excitations nor single-particle excitations across the gap. The peaks located at $(\pm\Delta_g,\Delta_g)$ and $(\pm\Delta_g,0)$ scale as $E_0^3$, which means that the excitation with energy $\Delta_g$ occurs through the absorption of single photons from the high-energy tail of the power-spectrum (Fig.~\ref{fig:power}). 
Only once the pulse energy approaches the gap size, there is a significant enhancement of the signal, in particular the R' one. While there are indications of resonances associated with multi-photon absorption at $\Omega<\Delta_g$, such contributions should be of higher order in $E_0$. Also, the results are similar in the dynamic and static Hartree calculations, which shows that excitations of the amplitude mode play no important role in the most prominent features.

Amplitude mode excitations however contribute to the very weak peak at $(\omega_\tau,\omega_t)=(0,\Delta_g)$, whose intensity scales as $E_0^5$ (when increasing the field strength from $E_0=0.02$ to $E_0=0.2$, the intensity of the signal increases from $1.6\times 10^{-5}$ to 1.3). Here, the intensity of the signal with dynamical Hartree term grows strongly as $\Omega$ approaches $\Delta_g$ (see  Fig.~\ref{fig:two_cycle_cut3}(e,f)). 
For $\Omega\gtrsim \Delta_g$, the first pulse can create a population state analogous to Fig.~\ref{fig:broadband}(b), and produce amplitude mode
excitations, which are then reabsorbed in subsequent pulses.
In this case, we also find an enhancement (relative to the static-Hartree calculation) at $\Omega=\Delta_g/2$, even though the signal amplitude is very weak.  

\begin{figure}[t]
	\centering
	\includegraphics[width=0.8\linewidth]{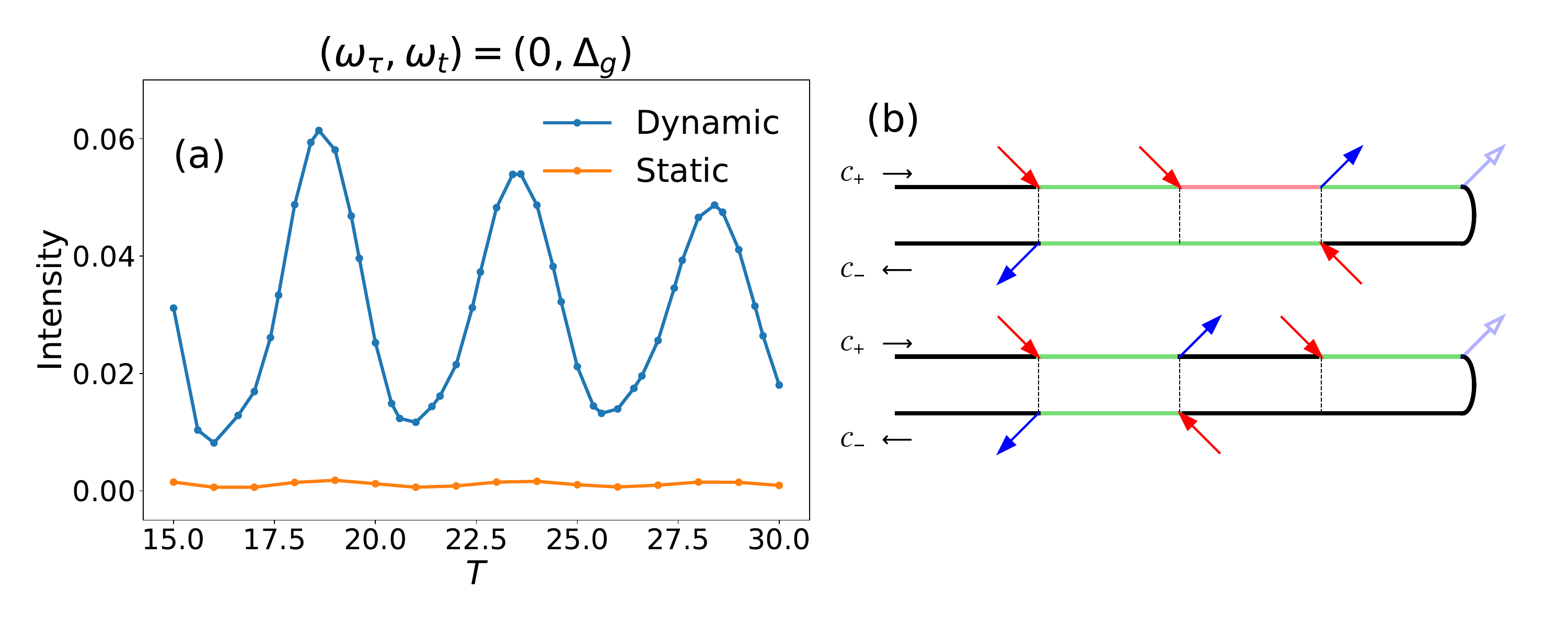}
	\caption{(a) Intensity of the signal at $(\omega_\tau,\omega_t)=(0,\Delta_g)$ as a function of the waiting time $T$. The dominant oscillation of the blue curve has a frequency $\sim\Delta_g$. The orange curve is calculated with the static Hartree term.
		(b) Two possible diagrams consistent with a $E_0^5$ scaling are presented. The top diagram leads to oscillations with frequency $\Delta_g$ as a function of the waiting time $T$. 
	}\label{fig:two_cycle3_intensity}
\end{figure}	

Looking at the waiting-time ($T$) dependence of this weak signal, shown in Fig.~\ref{fig:two_cycle3_intensity}(a), one finds a clear oscillation with
frequency $\Delta_g$ (period $2\pi/1.3=4.8$). A  diagram consistent with such $\Delta_g$ oscillations and the $E_0^5$ scaling is shown in the top subpanel of Fig.~\ref{fig:two_cycle3_intensity}(b).  
Other processes with a population state during the waiting time, like the one sketched in the bottom subpanel, may also contribute and produce an approximately constant shift in the intensity. If we repeat the calculations with static Hartree terms, the signal is strongly suppressed, which clearly shows that it is connected to amplitude mode excitations. 

Comparing the left and right panels in Fig.~\ref{fig:two_cycle_cut3} we see that the local potential disorder does not produce a qualitative change in the signals, but reduces their intensity. The perhaps most noteworthy observation is that in the case of the rephasing peak at $(\omega_\tau,\omega_t)=(-\Delta_g,\Delta_g)$ the signal from the dynamic Hartree simulation becomes smaller than the one from the static Hartree simulation in the presence of disorders. This means that processes involving amplitude mode excitations destructively interfere with those involving only single-particle excitations. 

\section{Discussions}
\label{sec:conclusions}

We analyzed the 2DCS signals of weak-coupling antiferromagnets using two-pulse and three-pulse protocols with broad and narrow power spectra. Since the staggered magnetization is encoded by the Hartree term, the comparison of simulations with static and dynamic Hartree term allowed us to identify the contribution of the amplitude mode of the antiferromagnet to the different features in the spectrum. This analysis in particular revealed that in the case of a clean system and for broadband pulses, the prominent rephasing and nonrephasing peaks at $(\omega_\tau,\omega_t)\approx (\pm \Delta_g, \Delta_g)$ 
are dominated by amplitude mode contributions, while the~R' ridge along $(\omega_\tau,\omega_t)\approx (-\Delta_g-x, x)$ originates from processes involving single-particle excitations and deexcitations. By combining this information with the waiting-time dependence of the signal intensities, we were able to sketch the relevant diagrams for the evolution of the density matrix. These diagrams provide insights into the time sequence of the excitation and deexcitation processes underlying the peaks in the 2DCS signals, and help to clarify the role of the amplitude mode excitations in these processes. 

We also studied the effect of local potential disorder on the 2DCS spectra and found that in contrast to superconductors, disorder does not play an essential role in antiferromagnets. Already in the clean system, there exist signals dominated by the amplitude mode contributions. While disorder weakens the signal intensity, it does not qualitatively change the relative contribution from single-particle and collective excitations. Weak-coupling antiferromagnets should thus be an ideal platform to study collective modes in nonlinear optical spectroscopy. 

In order to connect to the discussion of the amplitude mode resonance in the nonlinear response of superconductors \cite{tsuji2016, matsunaga2017}, we also studied the 2DCS signals measured with different narrowband (two-cycle) pulses. 
In the case of the two-pulse protocol, this analysis revealed a strong enhancement of the nonrephasing signal at $(\omega_\tau,\omega_t)=(\Delta_g,\Delta_g)$ for pulse frequency $\Omega\approx \Delta_g/2$, with a dominant contribution from amplitude mode related processes. In the three-pulse protocol, on the other hand, this resonance is absent, and the amplitude mode contribution instead dominates the (weak) signal at $(\omega_\tau,\omega_t)=(0,\Delta_g)$, which is associated with a fifth-order process.  

Finally, we have analyzed the power spectra of the nonlinear current using narrow-band pulses with fixed time delay, in close analogy to the recent THz 2DCS experiments for superconductors \cite{katsumi2024, katsumi2025}. We found that the first-harmonic response shows a sharp peak (as a function of the pulse frequency) at half of the gap energy, and that this signal is dominated by the amplitude mode contribution.
There is also a broad excitation continuum above the gap. 

Our results show that the 2DCS technique provides a useful tool for detecting collective modes in ordered phases. 
In this paper, we simulated collinear pulse setups in a minimal model for an antiferromagnet. More realistic simulations of materials and of non-collinear setups such as four-wave-mixing~\cite{Gelin2009} will require calculations with momentum-dependent Green's functions and possibly separate wave vectors of the pump pulses. 
In a future study, it will be interesting to apply such an analysis to superconducting states.

\section*{Acknowledgements}
J.C. and P.W. were supported by the Swiss National Science Foundation through the Research Unit QUAST of Deutsche Foschungsgemeinschaft (FOR5249).  The calculations were performed on the Beo05 cluster at the University of Fribourg, using a code based on \verb|NESSi|~\cite{schuler2020}. N.T. acknowledges support by JST FOREST (Grant No.~JPMJFR2131)
and JSPS KAKENHI (Grant Nos.~JP24H00191, JP25H01246, and JP25H01251).

\bibliography{ref}	

@article{tsuji2026,
  title = {Two-dimensional coherent spectroscopy of disordered superconductors in the narrow-band and broad-band limits},
  volume = {113},
  ISSN = {2469-9969},
  url = {http://dx.doi.org/10.1103/68kq-n6g8},
  DOI = {10.1103/68kq-n6g8},
  number = {21},
  journal = {Physical Review B},
  publisher = {American Physical Society (APS)},
  author = {Tsuji,  Naoto},
  year = {2026},
  month = June 
}

@misc{salvador2025,
  doi = {10.48550/arXiv.2501.16856},
  url = {https://arxiv.org/abs/2501.16856},
  author = {Salvador,  Alex Gómez and Morera,  Ivan and Michael,  Marios H. and Dolgirev,  Pavel E. and Pavicevic,  Danica and Liu,  Albert and Cavalleri,  Andrea and Demler,  Eugene},
  title = {Two-dimensional spectroscopy of bosonic collective excitations in disordered many-body systems},
  publisher = {arXiv},
  year = {2025},
}

@article{Gelin2009,
  title = {Efficient Calculation of Time- and Frequency-Resolved Four-Wave-Mixing Signals},
  volume = {42},
  ISSN = {1520-4898},
  url = {http://dx.doi.org/10.1021/ar900045d},
  DOI = {10.1021/ar900045d},
  number = {9},
  journal = {Accounts of Chemical Research},
  publisher = {American Chemical Society (ACS)},
  author = {Gelin,  Maxim F. and Egorova,  Dassia and Domcke,  Wolfgang},
  year = {2009},
  month = may,
  pages = {1290–1298}
}

@article{Shiba1972,
    author = {Shiba, Hiroyuki},
    title = {Thermodynamic Properties of the One-Dimensional Half-Filled-Band Hubbard Model. II: Application of the Grand Canonical Method},
    journal = {Progress of Theoretical Physics},
    volume = {48},
    number = {6},
    pages = {2171-2186},
    year = {1972},
    month = {12},
    issn = {0033-068X},
    doi = {10.1143/PTP.48.2171},
    url = {https://doi.org/10.1143/PTP.48.2171},
    eprint = {https://academic.oup.com/ptp/article-pdf/48/6/2171/5254982/48-6-2171.pdf},
}

@article{golez2020,
  title = {Nonlinear Spectroscopy of Collective Modes in an Excitonic Insulator},
  author = {Gole\ifmmode \check{z}\else \v{z}\fi{}, Denis and Sun, Zhiyuan and Murakami, Yuta and Georges, Antoine and Millis, Andrew J.},
  journal = {Phys. Rev. Lett.},
  volume = {125},
  issue = {25},
  pages = {257601},
  numpages = {6},
  year = {2020},
  month = {Dec},
  publisher = {American Physical Society},
  doi = {10.1103/PhysRevLett.125.257601},
  url = {https://link.aps.org/doi/10.1103/PhysRevLett.125.257601}
}

@article{pruschke2003,
doi = {10.1088/0953-8984/15/46/006},
url = {https://dx.doi.org/10.1088/0953-8984/15/46/006},
year = {2003},
month = {nov},
publisher = {},
volume = {15},
number = {46},
pages = {7867},
author = {Th. Pruschke and R. Zitzler},
title = {{From Slater to Mott–Heisenberg physics: the antiferromagnetic phase of the Hubbard model}},
journal = {J. Phys.: Condens. Matter}
}

@article{katsumi2025,
  title = {Amplitude Mode in a Multigap Superconductor $\mathrm{Mg}{\mathrm{B}}_{2}$ Investigated by Terahertz Two-Dimensional Coherent Spectroscopy},
  author = {Katsumi, Kota and Liang, Jiahao and Romero, Ralph and Chen, Ke and Xi, Xiaoxing and Armitage, N. P.},
  journal = {Phys. Rev. Lett.},
  volume = {135},
  issue = {3},
  pages = {036902},
  numpages = {5},
  year = {2025},
  month = {Jul},
  publisher = {American Physical Society},
  doi = {10.1103/g5rp-6vb1},
  url = {https://link.aps.org/doi/10.1103/g5rp-6vb1}
}

@article{katsumi2024,
  title = {Revealing Novel Aspects of Light-Matter Coupling by Terahertz Two-Dimensional Coherent Spectroscopy: The Case of the Amplitude Mode in Superconductors},
  author = {Katsumi, Kota and Fiore, Jacopo and Udina, Mattia and Romero, Ralph and Barbalas, David and Jesudasan, John and Raychaudhuri, Pratap and Seibold, Goetz and Benfatto, Lara and Armitage, N. P.},
  journal = {Phys. Rev. Lett.},
  volume = {132},
  issue = {25},
  pages = {256903},
  numpages = {7},
  year = {2024},
  month = {Jun},
  publisher = {American Physical Society},
  doi = {10.1103/PhysRevLett.132.256903},
  url = {https://link.aps.org/doi/10.1103/PhysRevLett.132.256903}
}

@article{matsunaga2014,
	title={{Light-induced collective pseudospin precession resonating with Higgs mode in a superconductor}},
	author={Matsunaga, Ryusuke and Tsuji, Naoto and Fujita, Hiroyuki and Sugioka, Arata and Makise, Kazumasa and Uzawa, Yoshinori and Terai, Hirotaka and Wang, Zhen and Aoki, Hideo and Shimano, Ryo},
	journal={Science},
	volume={345},
	number={6201},
	pages={1145--1149},
	year={2014},
	publisher={American Association for the Advancement of Science},
    doi = {10.1126/science.1254697},
    URL = {https://www.science.org/doi/abs/10.1126/science.1254697}
}

@article{tsuji2016,
  title = {{Nonlinear light--Higgs coupling in superconductors beyond BCS: Effects of the retarded phonon-mediated interaction}},
  author = {Tsuji, Naoto and Murakami, Yuta and Aoki, Hideo},
  journal = {Phys. Rev. B},
  volume = {94},
  issue = {22},
  pages = {224519},
  numpages = {13},
  year = {2016},
  month = {Dec},
  publisher = {American Physical Society},
  doi = {10.1103/PhysRevB.94.224519},
  url = {https://link.aps.org/doi/10.1103/PhysRevB.94.224519}
}

@article{katsumi2017,
  title = {{Higgs Mode in the $d$-Wave Superconductor ${\mathrm{Bi}}_{2}{\mathrm{Sr}}_{2}{\mathrm{CaCu}}_{2}{\mathrm{O}}_{8+x}$ Driven by an Intense Terahertz Pulse}},
  author = {Katsumi, Kota and Tsuji, Naoto and Hamada, Yuki I. and Matsunaga, Ryusuke and Schneeloch, John and Zhong, Ruidan D. and Gu, Genda D. and Aoki, Hideo and Gallais, Yann and Shimano, Ryo},
  journal = {Phys. Rev. Lett.},
  volume = {120},
  issue = {11},
  pages = {117001},
  numpages = {6},
  year = {2018},
  month = {Mar},
  publisher = {American Physical Society},
  doi = {10.1103/PhysRevLett.120.117001},
  url = {https://link.aps.org/doi/10.1103/PhysRevLett.120.117001}
}

@article{jujo2018,
author = {Jujo ,Takanobu},
title = {Quasiclassical Theory on Third-Harmonic Generation in Conventional Superconductors with Paramagnetic Impurities},
journal = {J. Phys. Soc. Jpn.},
volume = {87},
number = {2},
pages = {024704},
year = {2018},
doi = {10.7566/JPSJ.87.024704},
URL = {https://doi.org/10.7566/JPSJ.87.024704
}
}

@article{murotani2019,
  title = {Nonlinear optical response of collective modes in multiband superconductors assisted by nonmagnetic impurities},
  author = {Murotani, Yuta and Shimano, Ryo},
  journal = {Phys. Rev. B},
  volume = {99},
  issue = {22},
  pages = {224510},
  numpages = {15},
  year = {2019},
  month = {Jun},
  publisher = {American Physical Society},
  doi = {10.1103/PhysRevB.99.224510},
  url = {https://link.aps.org/doi/10.1103/PhysRevB.99.224510}
}

@article{silaev2019,
  title = {{Nonlinear electromagnetic response and Higgs-mode excitation in BCS superconductors with impurities}},
  author = {Silaev, Mikhail},
  journal = {Phys. Rev. B},
  volume = {99},
  issue = {22},
  pages = {224511},
  numpages = {16},
  year = {2019},
  month = {Jun},
  publisher = {American Physical Society},
  doi = {10.1103/PhysRevB.99.224511},
  url = {https://link.aps.org/doi/10.1103/PhysRevB.99.224511}
}

@article{schwarz2020,
	author = {Schwarz, L. and Fauseweh, B. and Tsuji, N. and Cheng, N. and Bittner, N. and Krull, H. and Berciu, M. and Uhrig, G. S. and Schnyder, A. P. and Kaiser, S. and Manske, D.},
	date = {2020/01/15},
	doi = {10.1038/s41467-019-13763-5},
	id = {Schwarz2020},
	isbn = {2041-1723},
	journal = {Nat. Commun.},
	number = {1},
	pages = {287},
	title = {{Classification and characterization of nonequilibrium Higgs modes in unconventional superconductors}},
	url = {https://doi.org/10.1038/s41467-019-13763-5},
	volume = {11},
	year = {2020}}

@article{shimano2020,
	title = {Higgs mode in superconductors},
	volume = {11},
	issn = {1947-5454, 1947-5462},
	url = {https://www.annualreviews.org/doi/10.1146/annurev-conmatphys-031119-050813},
	doi = {10.1146/annurev-conmatphys-031119-050813},
	number = {1},
	urldate = {2022-10-08},
	journal = {Annu. Rev. Condens. Matter Phys.},
	author = {Shimano, Ryo and Tsuji, Naoto},
	month = mar,
	year = {2020},
	pages = {103--124}
}

@article{baym1961,
  title = {Conservation Laws and Correlation Functions},
  author = {Baym, Gordon and Kadanoff, Leo P.},
  journal = {Phys. Rev.},
  volume = {124},
  issue = {2},
  pages = {287--299},
  numpages = {0},
  year = {1961},
  month = {Oct},
  publisher = {American Physical Society},
  doi = {10.1103/PhysRev.124.287},
  url = {https://link.aps.org/doi/10.1103/PhysRev.124.287}
}

@article{tsuji2015,
  title = {{Theory of Anderson pseudospin resonance with Higgs mode in superconductors}},
  author = {Tsuji, Naoto and Aoki, Hideo},
  journal = {Phys. Rev. B},
  volume = {92},
  issue = {6},
  pages = {064508},
  numpages = {11},
  year = {2015},
  month = {Aug},
  publisher = {American Physical Society},
  doi = {10.1103/PhysRevB.92.064508},
  url = {https://link.aps.org/doi/10.1103/PhysRevB.92.064508}
}

@article{matsunaga2017,
  title = {{Polarization-resolved terahertz third-harmonic generation in a single-crystal superconductor NbN: Dominance of the Higgs mode beyond the BCS approximation}},
  author = {Matsunaga, Ryusuke and Tsuji, Naoto and Makise, Kazumasa and Terai, Hirotaka and Aoki, Hideo and Shimano, Ryo},
  journal = {Phys. Rev. B},
  volume = {96},
  issue = {2},
  pages = {020505},
  numpages = {5},
  year = {2017},
  month = {Jul},
  publisher = {American Physical Society},
  doi = {10.1103/PhysRevB.96.020505},
  url = {https://link.aps.org/doi/10.1103/PhysRevB.96.020505}
}

@article{freericks2006,
  title = {Nonequilibrium Dynamical Mean-Field Theory},
  author = {Freericks, J. K. and Turkowski, V. M. and Zlati\ifmmode \acute{c}\else \'{c}\fi{}, V.},
  journal = {Phys. Rev. Lett.},
  volume = {97},
  issue = {26},
  pages = {266408},
  numpages = {4},
  year = {2006},
  month = {Dec},
  publisher = {American Physical Society},
  doi = {10.1103/PhysRevLett.97.266408},
  url = {https://link.aps.org/doi/10.1103/PhysRevLett.97.266408}
}

@article{tsuji2013,
  title = {Nonequilibrium dynamical mean-field theory based on weak-coupling perturbation expansions: Application to dynamical symmetry breaking in the Hubbard model},
  author = {Tsuji, Naoto and Werner, Philipp},
  journal = {Phys. Rev. B},
  volume = {88},
  issue = {16},
  pages = {165115},
  numpages = {28},
  year = {2013},
  month = {Oct},
  publisher = {American Physical Society},
  doi = {10.1103/PhysRevB.88.165115},
  url = {https://link.aps.org/doi/10.1103/PhysRevB.88.165115}
}

@article{schuler2020,
	title = {{NESSi}: {The} {Non}-{Equilibrium} {Systems} {Simulation} package},
	volume = {257},
	issn = {0010-4655},
	shorttitle = {{NESSi}},
	url = {https://www.sciencedirect.com/science/article/pii/S0010465520302277},
	doi = {10.1016/j.cpc.2020.107484},
	urldate = {2022-03-10},
	journal = {Comput. Phys. Commun.},
	author = {Schüler, Michael and Golež, Denis and Murakami, Yuta and Bittner, Nikolaj and Herrmann, Andreas and Strand, Hugo U. R. and Werner, Philipp and Eckstein, Martin},
	month = dec,
	year = {2020},
	pages = {107484},
}

@article{kemper2018,
  title = {General Principles for the Nonequilibrium Relaxation of Populations in Quantum Materials},
  author = {Kemper, A. F. and Abdurazakov, O. and Freericks, J. K.},
  journal = {Phys. Rev. X},
  volume = {8},
  issue = {4},
  pages = {041009},
  numpages = {14},
  year = {2018},
  month = {Oct},
  publisher = {American Physical Society},
  doi = {10.1103/PhysRevX.8.041009},
  url = {https://link.aps.org/doi/10.1103/PhysRevX.8.041009}
}

@article{endres2012,
  title = {The ‘Higgs’ amplitude mode at the two-dimensional superfluid/Mott insulator transition},
  volume = {487},
  ISSN = {1476-4687},
  url = {http://dx.doi.org/10.1038/nature11255},
  DOI = {10.1038/nature11255},
  number = {7408},
  journal = {Nature},
  publisher = {Springer Science and Business Media LLC},
  author = {Endres,  Manuel and Fukuhara,  Takeshi and Pekker,  David and Cheneau,  Marc and Schauβ,  Peter and Gross,  Christian and Demler,  Eugene and Kuhr,  Stefan and Bloch,  Immanuel},
  year = {2012},
  month = jul,
  pages = {454–458}
}

@article{tsuji2020,
  title = {{Higgs-mode resonance in third harmonic generation in NbN superconductors: Multiband electron-phonon coupling, impurity scattering, and polarization-angle dependence}},
  author = {Tsuji, Naoto and Nomura, Yusuke},
  journal = {Phys. Rev. Res.},
  volume = {2},
  issue = {4},
  pages = {043029},
  numpages = {16},
  year = {2020},
  month = {Oct},
  publisher = {American Physical Society},
  doi = {10.1103/PhysRevResearch.2.043029},
  url = {https://link.aps.org/doi/10.1103/PhysRevResearch.2.043029}
}

@article{chen2025,
  title = {Multidimensional coherent spectroscopy of correlated lattice systems},
  volume = {11},
  ISSN = {2057-3960},
  url = {http://dx.doi.org/10.1038/s41524-025-01619-0},
  DOI = {10.1038/s41524-025-01619-0},
  pages = {127},
  journal = {npj Computational Materials},
  author = {Chen,  Jiyu and Werner,  Philipp},
  year = {2025},
  month = may 
}

@book{mukamel1995,
	title = {Principles of {Nonlinear} {Optical} {Spectroscopy}},
	isbn = {978-0-19-509278-3},
	publisher = {Oxford University Press},
	author = {Mukamel, Shaul},
	year = {1995},
}

@book{hamm2011,
	address = {Cambridge},
	title = {Concepts and {Methods} of {2D} {Infrared} {Spectroscopy}},
	isbn = {978-1-107-00005-6},
	url = {https://www.cambridge.org/core/books/concepts-and-methods-of-2d-infrared-spectroscopy/8D35AA43C878AF1812CDCAF8890C9FE6},
	publisher = {Cambridge University Press},
	author = {Hamm, Peter and Zanni, Martin},
	year = {2011},
	doi = {10.1017/CBO9780511675935},
}

@misc{barbalas2023,
	title = "Energy {Relaxation} and dynamics in the correlated metal {Sr}$_2${RuO}$_4$ via {THz} two-dimensional coherent spectroscopy",
	url = {http://arxiv.org/abs/2312.13502},
	doi = {10.48550/arXiv.2312.13502},
	author = {Barbalas, David and Romero III, Ralph and Chaudhuri, Dipanjan and Mahmood, Fahad and Nair, Hari P. and Schreiber, Nathaniel J. and Schlom, Darrel G. and Shen, K. M. and Armitage, N. P.},
	month = dec,
	year = {2023},
}

@article{liu2025,
  title={Multidimensional terahertz probes of quantum materials},
  author={Liu, Albert},
  journal={npj Quantum Materials},
  volume={10},
  number={1},
  pages={18},
  year={2025},
  doi={10.1038/s41535-025-00741-y},
  url={https://www.nature.com/articles/s41535-025-00741-y},
  publisher={Nature Publishing Group UK London}
}

@article{georges1996,
  title = {Dynamical mean-field theory of strongly correlated fermion systems and the limit of infinite dimensions},
  author = {Georges, Antoine and Kotliar, Gabriel and Krauth, Werner and Rozenberg, Marcelo J.},
  journal = {Rev. Mod. Phys.},
  volume = {68},
  issue = {1},
  pages = {13--125},
  numpages = {0},
  year = {1996},
  month = {Jan},
  doi = {10.1103/RevModPhys.68.13},
  url = {https://link.aps.org/doi/10.1103/RevModPhys.68.13}
}

@article{boschini2024,
  title = {Time-resolved ARPES studies of quantum materials},
  author = {Boschini, Fabio and Zonno, Marta and Damascelli, Andrea},
  journal = {Rev. Mod. Phys.},
  volume = {96},
  issue = {1},
  pages = {015003},
  numpages = {56},
  year = {2024},
  month = {Feb},
  doi = {10.1103/RevModPhys.96.015003},
  url = {https://link.aps.org/doi/10.1103/RevModPhys.96.015003}
}

@article{zhang2024,
	title = {Terahertz-field-driven magnon upconversion in an antiferromagnet},
	issn = {1745-2481},
	url = {https://www.nature.com/articles/s41567-023-02350-7},
	doi = {10.1038/s41567-023-02350-7},
	journal = {Nat. Phys.},
	author = {Zhang, Zhuquan and Gao, Frank Y. and Chien, Yu-Che and Liu, Zi-Jie and Curtis, Jonathan B. and Sung, Eric R. and Ma, Xiaoxuan and Ren, Wei and Cao, Shixun and Narang, Prineha and von Hoegen, Alexander and Baldini, Edoardo and Nelson, Keith A.},
	month = jan,
    volume = {20},
	year = {2024},
	pages = {788}
}

@article{liu2024,
	title = {Probing inhomogeneous cuprate superconductivity by terahertz {Josephson} echo spectroscopy},
	issn = {1745-2481},
	url = {https://www.nature.com/articles/s41567-024-02643-5},
	doi = {10.1038/s41567-024-02643-5},
	journal = {Nat. Phys.},
	author = {Liu, A. and Pavićević, D. and Michael, M. H. and princ, A. G. and Dolgirev, P. E. and Fechner, M. and Disa, A. S. and M. Lozano, P. and Li, Q. and Gu, G. D. and Demler, E. and Cavalleri, A.},
	month = sep,
    volume = {20},
	year = {2024},
	pages = {1751}
}

@article{zhang2023,
	title = {Revealing the frequency-dependent oscillations in the nonlinear terahertz response induced by the {Josephson} current},
	volume = {10},
	issn = {2095-5138},
	url = {https://doi.org/10.1093/nsr/nwad163},
	doi = {10.1093/nsr/nwad163},
	number = {11},
	journal = {Natl. Sci. Rev.},
	author = {Zhang, Sijie and Sun, Zhiyuan and Liu, Qiaomei and Wang, Zixiao and Wu, Qiong and Yue, Li and Xu, Shuxiang and Hu, Tianchen and Li, Rongsheng and Zhou, Xinyu and Yuan, Jiayu and Gu, Genda and Dong, Tao and Wang, Nanlin},
	month = nov,
	year = {2023},
	pages = {nwad163},
}

@misc{chaudhuri2025,
	title = {Planckian dissipation, anomalous high temperature {THz} non-linear response and energy relaxation in the strange metal state of the cuprate superconductors},
	url = {http://arxiv.org/abs/2503.15646},
	doi = {10.48550/arXiv.2503.15646},
	urldate = {2025-04-05},
	author = {Chaudhuri, Dipanjan and Barbalas, David and Mahmood, Fahad and Liang, Jiahao and Romero III, Ralph and Legros, Anaelle and He, Xi and Raffy, Helene and Bozovic, Ivan and Armitage, N. P.},
	month = mar,
	year = {2025},
}

@article{lu2017,
  title = {Coherent Two-Dimensional Terahertz Magnetic Resonance Spectroscopy of Collective Spin Waves},
  author = {Lu, Jian and Li, Xian and Hwang, Harold Y. and Ofori-Okai, Benjamin K. and Kurihara, Takayuki and Suemoto, Tohru and Nelson, Keith A.},
  journal = {Phys. Rev. Lett.},
  volume = {118},
  issue = {20},
  pages = {207204},
  numpages = {6},
  year = {2017},
  month = {May},
  publisher = {American Physical Society},
  doi = {10.1103/PhysRevLett.118.207204},
  url = {https://link.aps.org/doi/10.1103/PhysRevLett.118.207204}
}

@article{folpini2017,
  title = {Strong Local-Field Enhancement of the Nonlinear Soft-Mode Response in a Molecular Crystal},
  author = {Folpini, Giulia and Reimann, Klaus and Woerner, Michael and Elsaesser, Thomas and Hoja, Johannes and Tkatchenko, Alexandre},
  journal = {Phys. Rev. Lett.},
  volume = {119},
  issue = {9},
  pages = {097404},
  numpages = {6},
  year = {2017},
  month = {Sep},
  publisher = {American Physical Society},
  doi = {10.1103/PhysRevLett.119.097404},
  url = {https://link.aps.org/doi/10.1103/PhysRevLett.119.097404}
}

@article{randi2017,
  title = {Bypassing the energy-time uncertainty in time-resolved photoemission},
  author = {Randi, Francesco and Fausti, Daniele and Eckstein, Martin},
  journal = {Phys. Rev. B},
  volume = {95},
  issue = {11},
  pages = {115132},
  numpages = {11},
  year = {2017},
  month = {Mar},
  publisher = {American Physical Society},
  doi = {10.1103/PhysRevB.95.115132},
  url = {https://link.aps.org/doi/10.1103/PhysRevB.95.115132}
}

@misc{fiore2025,
  doi = {10.48550/arXiv.2509.25060},
  url = {https://arxiv.org/abs/2509.25060},
  author = {Fiore,  Jacopo and Sellati,  Niccolò and Udina,  Mattia and Benfatto,  Lara},
  title = {Two-dimensional THz spectroscopy in electronic systems: a many-body diagrammatic approach},
  year = {2025},
}

@misc{gomez2025,
  title = {Gap Inhomogeneity in Cuprates: a view from Two-Dimensional Josephson Echo Spectroscopy},
  doi = {10.48550/arXiv.2509.23856},
  url = {https://arxiv.org/abs/2509.23856},
  author = {Gómez Salvador, Alex and Morera,  Ivan and Michael,  Marios H. and Dolgirev,  Pavel E. and Pavicevic,  Danica and Liu,  Albert and Cavalleri,  Andrea and Demler,  Eugene},
  year = {2025},
}

@article{gomez2024,
	title = {Principles of two-dimensional terahertz spectroscopy of collective excitations: {The} case of {Josephson} plasmons in layered superconductors},
	volume = {110},
	url = {https://link.aps.org/doi/10.1103/PhysRevB.110.094514},
	doi = {10.1103/PhysRevB.110.094514},
	number = {9},
	journal = {Phys. Rev. B},
	author = {Gómez Salvador, Alex and Dolgirev, Pavel E. and Michael, Marios H. and Liu, Albert and Pavicevic, Danica and Fechner, Michael and Cavalleri, Andrea and Demler, Eugene},
	month = sep,
	year = {2024},
	pages = {094514},
}

@article{li2024,
	title = {Time-domain interferometry of electron weak localization through terahertz nonlinear response},
	volume = {6},
	url = {https://link.aps.org/doi/10.1103/PhysRevResearch.6.033125},
	doi = {10.1103/PhysRevResearch.6.033125},
	number = {3},
	journal = {Physical Review Research},
	author = {Li, Zi-Long and Li, Xiao-Hui and Wan, Yuan},
	month = aug,
	year = {2024},
	pages = {033125},
}

@article{li2023,
	title = {Photon echo and fractional excitation lensing of the {S}$=\frac{1}{2}$ {XY} spin chain},
	volume = {108},
	issn = {2469-9950, 2469-9969},
	url = {https://link.aps.org/doi/10.1103/PhysRevB.108.165151},
	doi = {10.1103/PhysRevB.108.165151},
	number = {16},
	journal = {Physical Review B},
	author = {Li, Zi-Long and Wan, Yuan},
	month = oct,
	year = {2023},
	pages = {165151},
}

@article{gao2023,
	title = {Two-dimensional coherent spectrum of interacting spinons from matrix product states},
	volume = {107},
	url = {https://link.aps.org/doi/10.1103/PhysRevB.107.165121},
	doi = {10.1103/PhysRevB.107.165121},
	number = {16},
	journal = {Phys. Rev. B},
	author = {Gao, Qi and Liu, Yang and Liao, Haijun and Wan, Yuan},
	month = apr,
	year = {2023},
	pages = {165121},
}

@article{li2021,
	title = {Photon {Echo} from {Lensing} of {Fractional} {Excitations} in {Tomonaga}-{Luttinger} {Spin} {Liquid}},
	volume = {11},
	issn = {2160-3308},
	url = {https://link.aps.org/doi/10.1103/PhysRevX.11.031035},
	doi = {10.1103/PhysRevX.11.031035},
	number = {3},
	journal = {Phys. Rev. X},
	author = {Li, Zi-Long and Oshikawa, Masaki and Wan, Yuan},
	month = aug,
	year = {2021},
	pages = {031035},
}

@article{wan2019,
	title = {Resolving {Continua} of {Fractional} {Excitations} by {Spinon} {Echo} in {THz} {2D} {Coherent} {Spectroscopy}},
	volume = {122},
	url = {https://link.aps.org/doi/10.1103/PhysRevLett.122.257401},
	doi = {10.1103/PhysRevLett.122.257401},
	number = {25},
	journal = {Phys. Rev. Lett.},
	author = {Wan, Yuan and Armitage, N. P.},
	month = jun,
	year = {2019},
	pages = {257401},
}

@article{branczyk2014,
	title = {Crossing disciplines - {A} view on two-dimensional optical spectroscopy},
	volume = {526},
	issn = {1521-3889},
	url = {https://onlinelibrary.wiley.com/doi/abs/10.1002/andp.201300153},
	doi = {10.1002/andp.201300153},
	number = {1-2},
	journal = {Annalen der Physik},
	author = {Brańczyk, Agata M. and Turner, Daniel B. and Scholes, Gregory D.},
	year = {2014},
	pages = {31--49},
}

@article{aoki2014,
  title = {Nonequilibrium Dynamical Mean-Field Theory and Its Applications},
  author = {Aoki, Hideo and Tsuji, Naoto and Eckstein, Martin and Kollar, Marcus and Oka, Takashi and Werner, Philipp},
  year = {2014},
  month = jun,
  journal = {Rev. Mod. Phys.},
  volume = {86},
  number = {2},
  pages = {779--837},
  doi = {10.1103/RevModPhys.86.779},
  url = {https://link.aps.org/doi/10.1103/RevModPhys.86.779}
}

@article{werner2017,
	title = {{Ultrafast switching of composite order in  ${A}_{3}{\mathrm{C}}_{60}$}},
	volume = {95},
	url = {https://link.aps.org/doi/10.1103/PhysRevB.95.195405},
	doi = {10.1103/PhysRevB.95.195405},
	number = {19},
	journal = {Phys. Rev. B},
	author = {Werner, Philipp and Strand, Hugo U. R. and Hoshino, Shintaro and Eckstein, Martin},
	month = may,
	year = {2017},
	pages = {195405},
}

@article{eckstein2010,
  title = {Nonequilibrium dynamical mean-field calculations based on the noncrossing approximation and its generalizations},
  author = {Eckstein, Martin and Werner, Philipp},
  journal = {Phys. Rev. B},
  volume = {82},
  issue = {11},
  pages = {115115},
  numpages = {13},
  year = {2010},
  month = {Sep},
  doi = {10.1103/PhysRevB.82.115115},
  url = {https://link.aps.org/doi/10.1103/PhysRevB.82.115115}
}

@incollection{tsuji2024,
title = {{Higgs and Nambu–Goldstone modes in condensed matter physics}},
booktitle = {Encyclopedia of Condensed Matter Physics (Second Edition)},
publisher = {Academic Press},
address = {Oxford},
pages = {174-186},
year = {2024},
isbn = {978-0-323-91408-6},
doi = {https://doi.org/10.1016/B978-0-323-90800-9.00256-0},
url = {https://www.sciencedirect.com/science/article/pii/B9780323908009002560},
author = {Naoto Tsuji and Ippei Danshita and Shunji Tsuchiya}
}

@article{ruegg2008,
  title = {{Quantum Magnets under Pressure: Controlling Elementary Excitations in ${\mathrm{TlCuCl}}_{3}$}},
  author = {R\"uegg, Ch. and Normand, B. and Matsumoto, M. and Furrer, A. and McMorrow, D. F. and Kr\"amer, K. W. and G\"udel, H. -U. and Gvasaliya, S. N. and Mutka, H. and Boehm, M.},
  journal = {Phys. Rev. Lett.},
  volume = {100},
  issue = {20},
  pages = {205701},
  numpages = {4},
  year = {2008},
  month = {May},
  publisher = {American Physical Society},
  doi = {10.1103/PhysRevLett.100.205701},
  url = {https://link.aps.org/doi/10.1103/PhysRevLett.100.205701}
}

@article{sooryakumar1980,
  title = {{Raman Scattering by Superconducting-Gap Excitations and Their Coupling to Charge-Density Waves}},
  author = {Sooryakumar, R. and Klein, M. V.},
  journal = {Phys. Rev. Lett.},
  volume = {45},
  issue = {8},
  pages = {660--662},
  numpages = {0},
  year = {1980},
  month = {Aug},
  publisher = {American Physical Society},
  doi = {10.1103/PhysRevLett.45.660},
  url = {https://link.aps.org/doi/10.1103/PhysRevLett.45.660}
}

@article{grasset2018,
  title = {{Higgs-mode radiance and charge-density-wave order in $2H-{\mathrm{NbSe}}_{2}$}},
  author = {Grasset, Romain and Cea, Tommaso and Gallais, Yann and Cazayous, Maximilien and Sacuto, Alain and Cario, Laurent and Benfatto, Lara and M\'easson, Marie-Aude},
  journal = {Phys. Rev. B},
  volume = {97},
  issue = {9},
  pages = {094502},
  numpages = {12},
  year = {2018},
  month = {Mar},
  publisher = {American Physical Society},
  doi = {10.1103/PhysRevB.97.094502},
  url = {https://link.aps.org/doi/10.1103/PhysRevB.97.094502}
}

@article{chu2020,
	author = {Chu, Hao and Kim, Min-Jae and Katsumi, Kota and Kovalev, Sergey and Dawson, Robert David and Schwarz, Lukas and Yoshikawa, Naotaka and Kim, Gideok and Putzky, Daniel and Li, Zhi Zhong and Raffy, H{\'e}l{\`e}ne and Germanskiy, Semyon and Deinert, Jan-Christoph and Awari, Nilesh and Ilyakov, Igor and Green, Bertram and Chen, Min and Bawatna, Mohammed and Cristiani, Georg and Logvenov, Gennady and Gallais, Yann and Boris, Alexander V. and Keimer, Bernhard and Schnyder, Andreas P. and Manske, Dirk and Gensch, Michael and Wang, Zhe and Shimano, Ryo and Kaiser, Stefan},
	doi = {10.1038/s41467-020-15613-1},
	isbn = {2041-1723},
	journal = {Nat. Commun.},
	number = {1},
	pages = {1793},
	title = {{Phase-resolved Higgs response in superconducting cuprates}},
	url = {https://doi.org/10.1038/s41467-020-15613-1},
	volume = {11},
	year = {2020},
}

@article{matsunaga2013,
  title = {{Higgs Amplitude Mode in the BCS Superconductors ${\mathrm{Nb}}_{1\mathrm{\text{\ensuremath{-}}}x}{\mathrm{Ti}}_{x}\mathbf{N}$ Induced by Terahertz Pulse Excitation}},
  author = {Matsunaga, Ryusuke and Hamada, Yuki I. and Makise, Kazumasa and Uzawa, Yoshinori and Terai, Hirotaka and Wang, Zhen and Shimano, Ryo},
  journal = {Phys. Rev. Lett.},
  volume = {111},
  issue = {5},
  pages = {057002},
  numpages = {5},
  year = {2013},
  month = {Jul},
  publisher = {American Physical Society},
  doi = {10.1103/PhysRevLett.111.057002},
  url = {https://link.aps.org/doi/10.1103/PhysRevLett.111.057002}
}

@article{anderson1959,
title = {Theory of dirty superconductors},
journal = {J. Phys. Chem. Solids},
volume = {11},
number = {1},
pages = {26-30},
year = {1959},
issn = {0022-3697},
doi = {https://doi.org/10.1016/0022-3697(59)90036-8},
url = {https://www.sciencedirect.com/science/article/pii/0022369759900368},
author = {P.W. Anderson},
}

@article{cea2016,
  title = {{Nonlinear optical effects and third-harmonic generation in superconductors: Cooper pairs versus Higgs mode contribution}},
  author = {Cea, T. and Castellani, C. and Benfatto, L.},
  journal = {Phys. Rev. B},
  volume = {93},
  issue = {18},
  pages = {180507},
  numpages = {5},
  year = {2016},
  month = {May},
  publisher = {American Physical Society},
  doi = {10.1103/PhysRevB.93.180507},
  url = {https://link.aps.org/doi/10.1103/PhysRevB.93.180507}
}

@article{Seibold2021,
  title = {Third harmonic generation from collective modes in disordered superconductors},
  author = {Seibold, G. and Udina, M. and Castellani, C. and Benfatto, L.},
  journal = {Phys. Rev. B},
  volume = {103},
  issue = {1},
  pages = {014512},
  numpages = {19},
  year = {2021},
  month = {Jan},
  publisher = {American Physical Society},
  doi = {10.1103/PhysRevB.103.014512},
  url = {https://link.aps.org/doi/10.1103/PhysRevB.103.014512}
}

@article{Hong2017,
	author = {Hong, Tao and Matsumoto, Masashige and Qiu, Yiming and Chen, Wangchun and Gentile, Thomas R. and Watson, Shannon and Awwadi, Firas F. and Turnbull, Mark M. and Dissanayake, Sachith E. and Agrawal, Harish and Toft-Petersen, Rasmus and Klemke, Bastian and Coester, Kris and Schmidt, Kai P. and Tennant, David A.},
	date = {2017/07/01},
	doi = {10.1038/nphys4182},
	id = {Hong2017},
	isbn = {1745-2481},
	journal = {Nat. Phys.},
	number = {7},
	pages = {638--642},
	title = {Higgs amplitude mode in a two-dimensional quantum antiferromagnet near the quantum critical point},
	url = {https://doi.org/10.1038/nphys4182},
	volume = {13},
	year = {2017}}

@article{Jain2017,
	author = {Jain, A. and Krautloher, M. and Porras, J. and Ryu, G. H. and Chen, D. P. and Abernathy, D. L. and Park, J. T. and Ivanov, A. and Chaloupka, J. and Khaliullin, G. and Keimer, B. and Kim, B. J.},
	date = {2017/07/01},
	doi = {10.1038/nphys4077},
	id = {Jain2017},
	isbn = {1745-2481},
	journal = {Nat. Phys.},
	number = {7},
	pages = {633--637},
	title = {Higgs mode and its decay in a two-dimensional antiferromagnet},
	url = {https://doi.org/10.1038/nphys4077},
	volume = {13},
	year = {2017}}

@article{Pekker2015,
   author = "Pekker, David and Varma, C. M.",
   title = {{Amplitude/Higgs Modes in Condensed Matter Physics}}, 
   journal= "Annu. Rev. Condens. Matter Phys.",
   year = "2015",
   volume = "6",
   number = "Volume 6, 2015",
   pages = "269-297",
   doi = "https://doi.org/10.1146/annurev-conmatphys-031214-014350",
   url = "https://www.annualreviews.org/content/journals/10.1146/annurev-conmatphys-031214-014350",
   publisher = "Annual Reviews",
   issn = "1947-5462",
   type = "Journal Article"
  }

@article{Grasset2019,
  title = {{Pressure-Induced Collapse of the Charge Density Wave and Higgs Mode Visibility in $2H-{\mathrm{TaS}}_{2}$}},
  author = {Grasset, Romain and Gallais, Yann and Sacuto, Alain and Cazayous, Maximilien and Ma\~nas-Valero, Samuel and Coronado, Eugenio and M\'easson, Marie-Aude},
  journal = {Phys. Rev. Lett.},
  volume = {122},
  issue = {12},
  pages = {127001},
  numpages = {6},
  year = {2019},
  month = {Mar},
  publisher = {American Physical Society},
  doi = {10.1103/PhysRevLett.122.127001},
  url = {https://link.aps.org/doi/10.1103/PhysRevLett.122.127001}
}

@book{AbrikosovBook,
  title     = {{Methods of Quantum Field Theory in Statistical Physics}},
  author    = {Abrikosov, A. A. and Gorkov, L. P. and Dzyaloshinski, I. E.},
  publisher = {Dover Publications},
  address   = {New York},
  year      = {1975},
  isbn      = {978-0486632285}
}

\end{document}